\documentclass[prd,preprintnumbers,nofootinbib]{revtex4} 
\usepackage{graphicx} 
\usepackage{amsmath}
\usepackage{amsfonts,amsbsy}
\usepackage{amssymb}
\usepackage{appendix}
\usepackage{array}
\usepackage{multirow}
\def\be{\begin{equation}}
\def\ee{\end{equation}}
\def\bea{\begin{eqnarray}}
\def\eea{\end{eqnarray}}

\def\eq#1{{Eq.~(\ref{#1})}}
\def\fig#1{{Fig.~\ref{#1}}}

\def\beq{\begin{equation}}
\def\eeq{\end{equation}}
\def\bea{\begin{eqnarray}}
\def\eea{\end{eqnarray}}
\newcommand{\Lb}{\left(}
\newcommand{\Rb}{\right)}

\newlength{\dinwidth}
\newlength{\dinmargin}
\renewcommand{\vec}[1]{\boldsymbol{#1}}
\newcommand{\dif}{\mathrm{d}}

\newcommand{\chisq}{\chi^2/\mathrm{d.o.f.}}

\makeatother

\makeatother

\begin{document}

\title{Impact-parameter dependent Color Glass Condensate dipole model and new combined HERA data}

\author{Amir H. Rezaeian and Ivan Schmidt}
\affiliation{
 Departamento de F\'\i sica, Universidad T\'ecnica
Federico Santa Mar\'\i a, Avda. Espa\~na 1680,
Casilla 110-V, Valparaiso, Chile\\
Centro Cient\'\i fico Tecnol\'ogico de Valpara\'\i so (CCTVal), Universidad T\'ecnica
Federico Santa Mar\'\i a, Valpara\'\i so, Chile}
\date{\today}
\begin{abstract}
The Impact-Parameter dependent Color Glass Condensate (b-CGC) dipole model is based on the Balitsky-Kovchegov non-linear evolution equation and improves the Iancu-Itakura-Munier dipole model by incorporating the impact-parameter dependence of the saturation scale. Here we confront  the model to the recently released high precision combined HERA data and obtain its parameters.  The b-CGC results are then compared to data at small-x for the structure function, the longitudinal structure function, the charm structure function, exclusive vector meson ($J/\psi, \phi$ and $\rho$) production and Deeply Virtual Compton Scattering (DVCS). We also compare our results with the Impact-Parameter dependent Saturation model (IP-Sat). We show that most features of inclusive DIS and exclusive diffractive data, including the $Q^2$, $W$, $|t|$ and $x$ dependence are correctly reproduced in both models.  Nevertheless, the b-CGC and the IP-Sat models give different predictions beyond the current HERA kinematics, namely for the structure functions at very low $x$ and high virtualities $Q^2$, and for the exclusive diffractive vector meson and DVCS  production at  high $t$. This can be traced back to the different power-law behavior of the saturation scale in $x$, and to a different impact-parameter $b$ dependence  of the saturation scale in these models. Nevertheless, both models  give approximately similar saturation scale $Q_S<1$ GeV for the proton in HERA kinematics, and also  both models lead to the same conclusion that the typical impact-parameter probed in the total $\gamma^{*}p$ cross-section is about $b\approx 2\div 3\,\text{GeV}^{-1}$. Our results provide a benchmark for further investigation of QCD  at small-x in heavy ion collisions at RHIC and the LHC and also at future experiments such as an Electron-Ion Collider and the LHeC.

\end{abstract}
\maketitle

\section{Introduction}

It is believed that the experimental data in Deep inelastic scattering (DIS)  and exclusive diffractive processes in electron-proton collisions, such as exclusive vector meson production and deeply virtual Compton scattering  (DVCS) at small-x, can provide valuable information about the unitarity limits of QCD and  parton saturation effects  \cite{sg,mv,cgc-review1}.  Moreover, the small-x physics has important implications about our understanding of the early stages of relativistic heavy ion collisions at both RHIC and the LHC, where because of high field strengths and high density of colliding nuclei, one expects that the physics of parton saturation be important. Studies along  this line also provide  crucial benchmarks for further investigations of the high-energy limit of QCD at the LHC, and  also at future experiments such as an Electron-Ion Collider (EIC) \cite{eic} and the large Hadron Electron Collider (LHeC) \cite{lhec}. 

An effective perturbative weak-coupling field theory approach that can describe the small-x regime of QCD is the Color Glass Condensate (CGC)~\cite{mv,cgc-review1}.  In this formalism quantum corrections are systematically re-summed,  getting enhanced by large logarithms of $1/x$ and also incorporating high gluon density effects at small x and for large nuclei. The most important ingredient for particle production in the CGC approach is the universal dipole amplitude \cite{dipole1}, given by the imaginary part of the quark-antiquark scattering amplitude on a proton (or nuclear target). The rapidity evolution of this dipole amplitude, given a suitable initial condition, can be obtained by solving the Jalilian-Marian-Iancu-McLerran-Weigert-Leonidov-Kovner (JIMWLK)  hierarchy of equations~\cite{jimwlk} or in the large $N_c$ limit,  the Balitsky-Kovchegov (BK) equation \cite{bk,bb}.  

Unfortunately, impact-parameter dependent numerical solutions to these non-linear  JIMWLK and the BK equations are  
very difficult to obtain \cite{ana}. Moreover, the choice of the impact-parameter profile of the dipole amplitude entails intrinsically non-perturbative physics,  which is beyond the QCD weak-coupling approach to the JIMWLK and the BK equations. In fact, both of these small $x$ evolution equations generate a power law Coulomb-like tail, which is not confining at large distances \cite{al,bk-c,ana} and therefore may violate the unitarity bound. 
For these reasons, in practice a specific form of the impact parameter dependence of the dipole amplitude is assumed, and  the simplest choice is to take it to be constant.  Nevertheless, this approximation is not reliable since the gluon density is larger at the proton center than at its periphery, and in fact the saturation scale at the center of proton ($b=0$) can be about $2\div 4$ times larger than the saturation scale at the typical impact-parameter $b\approx 1-4~\text{GeV}^{-1}$ probed in DIS (see Sec. III). Since the  b-dependence of the dipole amplitude is essential for understanding exclusive diffractive processes in the CGC or the color dipole approach,  diffractive data at HERA provide extra valuable constrains on saturation models, which are inaccessible in b-independent dipole models.

A simple dipole model that incorporates all known properties of the gluon saturation in both the BK and the JIMWLK frameworks, and which models the impact parameter dependence of  dipole amplitude, is the b-CGC dipole model \cite{watt-bcgc,ip-sat1}.  This is actually an improved form of the Iancu-Itakura-Munier dipole model \cite{IIM} (the so-called CGC dipole model),  introducing the impact-parameter dependence of the saturation scale. In both the CGC and the b-CGC dipole model,  two well-known limiting regimes are matched, the one of the BFKL equation and the region deep inside the saturation, by simple analytical interpolations. There is also another well-known impact-parameter dependent saturation model, the so-called IP-Sat model \cite{ip-sat0,ip-sat2}. In this case the saturation boundary is approached via the DGLAP evolution, that is by the eikonalization of the gluon distribution, which effectively represents higher twist contributions. The b-CGC and the IP-Sat models have been both applied to various reactions, from DIS and diffractive processes \cite{watt-bcgc,IIM,ip-sat0,ip-sat1,ip-sat2,ip-g1,ip-g2} to proton-proton \cite{pp-LR,pp0,pp1}, proton-nucleus \cite{pa-R,Tribedy:2011aa,all-pa}, and nucleus-nucleus collisions \cite{aa-LR,Schenke:2012wb,lp-aa,aa-g1,aa-g2,aa-g3,aa-g4} at RHIC and the LHC. The difference between the b-CGC and the IP-Sat models comes in part from our current theoretical uncertainties in modeling the impact-parameter dependence in the dipole approach.

Recently, the H1 and ZEUS collaborations have released new combined data for inclusive DIS~\cite{Aaron:2009aa,Abramowicz:1900rp}. 
These data have extremely small error bars, and it is therefore vital to confront it with the b-CGC dipole mode, in order to examine the effects of the tighter constraints on model parameters. As a by-product, we will also reexamine the CGC dipole model in view of  recent precise data from HERA and obtain its free parameters from a fit.  Since the IP-Sat dipole model was also recently updated with the recent combined data from HERA \cite{ip-sat2}, we also compare the b-CGC and the IP-Sat results for both DIS and exclusive diffractive data at HERA, and provide predictions for various observable for a wide range of kinematics.

This paper is organized as follows. In section II, we introduce the main formulation of the color dipole approach for calculating the total DIS cross-section, structure functions and exclusive diffractive processes, discussing also the CGC and the b-CGC dipole models. In section III, we present a detailed numerical analysis and our main results. We then confront these results with HERA data for inclusive DIS and exclusive diffractive processes, in both the b-CGC and the IP-Sat models, and provide predictions for future experiments. We summarize our main results in section IV.

\section{\bf  Dipole description of DIS: a unified description of inclusive and exclusive diffractive processes}
\subsection{Proton structure functions and total DIS cross-section}
The proton structure function $F_2$ and the longitudinal structure function $F_L$, for a given Bjorken  $x$ and virtuality $Q^2$, can be written in terms of the
$\gamma^{\star}p$ total cross section as follows,
\begin{eqnarray}
F_2(Q^2,x) &=& \frac{Q^2}{4\pi^2\alpha_{EM}} 
\left[\sigma_L^{\gamma^*p}(Q^2,x)+\sigma_T^{\gamma^*p}(Q^2,x)\right],\label{f2}\\
F_L(Q^2,x) &=& \frac{Q^2}{4\pi^2\alpha_{EM}}\sigma_L^{\gamma^*p}(Q^2,x), \ \label{FL}
\end{eqnarray}
where $\alpha_{EM}$ is the electromagnetic fine structure constant and  the subscript $L,T$ denote the
longitudinal and transverse polarizations of the virtual photon. In the dipole picture, the scattering of the virtual photon $\gamma^{\star}$ on the proton can be conceived as a $\gamma^{\star}$ fluctuation into a quark-antiquark pair with size $r$ (the color dipole is flavor blind), in which the produced $q\bar{q}$ dipole then interacts with the proton via gluon exchanges. The lifetime of the $q\bar{q}$ dipole at small $x$ is much longer than its typical interaction time with the target. Therefore, the total deep inelastic cross-section  can be factorized  in the following form \cite{ni,al1},  
\begin{equation}\label{gp}
  \sigma_{L,T}^{\gamma^*p}(Q^2,x) = 2\sum_f \int\!\dif^2\vec{r}\int\!\dif^2\vec{b}\int_0^1 dz\, |\Psi_{L,T}^{(f)}(r,z,m_f;Q^2)|^2 
  \,\mathcal{N}\left(x,r,b\right),
\end{equation}
where $z$ is the fraction of the light-front momentum of the virtual
photon carried by the quark, and $m_f$ is the quark mass. In \eq{gp}, $\mathcal{N}\left(x,r,b\right)$ is the imaginary part of the forward $q\bar{q}$ dipole-proton scattering amplitude with dipole transverse-size $r$ and collision impact parameter $b$. The first part of the process, $\gamma^{\star}$ splitting to $q\bar{q}$ dipole, can be mainly described by QED while the later stage encoded in the dipole amplitude $\mathcal{N}(x, r, b)$ requires physics beyond the standard perturbative QCD,  and incorporates higher order gluon emissions and also the impact-parameter dependence of the collision (see below). The explicit form of light front wave function $\Psi_{L,T}^{(f)}$, for $\gamma^{\star}$ fluctuations into $q\bar{q}$ at the lowest order in $\alpha_{EM}$, can be found in Ref.\,\cite{gbw}.  

For the light quarks, the gluon density is evaluated at $x = x_{Bj}$ (Bjorken-x), while for charm quarks we take $x = x_{Bj} \, (1 + 4\,m_c^2/Q^2)$ \cite{gbw}.  The contribution of the charm quark  in the flavor summation  of wave functions in Eq.~(\ref{gp}),  directly gives the charm structure function $F_2^{c\bar{c}}$ via \eq{f2}. 

The reduced cross-section $\sigma_{r}$ is expressed in terms of the inclusive proton structure functions $F_2$ in \eq{f2} and  $F_L$ in \eq{FL}, 
\[
\sigma_{r}\left(Q^{2},x,y\right)=F_{2}\left(Q^{2},x\right)-\frac{y^{2}}{1+(1-y)^{2}}F_{L}\left(Q^{2},x\right),
\]
where $y=Q^2/(sx)$ is the inelasticity variable and $\sqrt{s}$ denotes the center of mass energy in $ep$ collisions\footnote{Here, we neglect the contribution of the $Z$ boson, which can become important only at very large $Q^2$.}. The advantage of the reduced cross-section $\sigma_r$ is that it is unbiased towards any theoretical assumption in the extraction of the structure functions $F_2$ and $F_L$. 


\subsection{Impact-parameter dependent color glass condensate dipole model  }
The common ingredient of the total (and reduced) cross-sections, proton structure functions in DIS, exclusive diffractive vector meson production and  DVCS, is the universal $q\bar{q}$ dipole-proton forward amplitude.  As we will show in the following sections, the impact-parameter dependence of the dipole amplitude is crucial for describing  exclusive diffractive processes. However, the general practice is that for the total cross-section in DIS as well as proton structure functions, the effect of the impact-parameter dependence of the dipole amplitude is ignored and it is effectively incorporated by treating it as an overall normalization. In this way, one can still find a good fit for the structure functions and total DIS cross-section. However, such a trivial $b$-dependence leads in general to a pronounced precocious dip in the $t$-distribution of vector meson production at rather low or moderate $|t|$, which is not supported by the available experimental data from HERA \cite{ip-sat0,ip-sat1}.

Iancu, Itakura and Munier \cite{IIM} proposed a simple dipole model, constructed by smoothly interpolating between two limiting behaviors which are analytically under control, namely the solution to the BFKL equation in the vicinity of the saturation line for small dipole sizes, $r <<1/Q_s$, and the Levin-Tuchin solution \cite{LT} of the BK equation deep inside the saturation region for larger dipoles, $r >>1/Q_s$ (see also Refs.\,\cite{lt0,lt1,lt2,lt3}). This model is historically called CGC dipole model. Notice that it was recently numerically shown that the JIMWLK equation indeed leads, to high accuracy, to the Levin-Tuchin formula  for the S-matrix close to the unitarity limit \cite{lt3}.  In the CGC dipole model, the color dipole-proton amplitude is given by, 
\bea \label{CA5}
N\Lb x, r, b\Rb\,\,=\,\, \left\{\begin{array}{l}\,\,\,N_0\,\Lb \frac{r Q_s}{2}\Rb^{2\gamma_{eff}}\,\,\,\,\,\,\,\,r Q_s\,\leq\,2\,,\\ \\
1\,\,-\,\,\exp\Lb -\mathcal{A} \ln^2\Lb \mathcal{B} r Q_s\Rb\Rb\,\,\,\,\,\,\,\,\,\,\,\,\,\,\,\ rQ_s\,>\,2\,,\end{array}
\right.
\eea
with effective anomalous dimension defined as 
\beq \label{g-eff}
\gamma_{eff}=\gamma_s\,\,+\,\,\frac{1}{\kappa \lambda Y}\ln\Lb\frac{2}{r Q_s}\Rb,
\end{equation}
where  $Y=\ln(1/x)$ and $\kappa= \chi''(\gamma_s)/\chi'(\gamma_s)$, with $\chi$ being the LO BFKL
characteristic function. The second term (diffusion term)  in $\gamma_{eff}$ enhances the anomalous dimension from its value at BFKL $\gamma_{eff}\to\gamma_s$ to DGLAP  $\gamma_{eff}\to  1$, matching the BFKL region to the color-transparency regime of the DGLAP for small dipole sizes\footnote{Notice that the anomalous dimension defined via \eq{g-eff} is not well-defined as $r\to 0$. However, this limiting case has negligible contribution to the total cross-section.} (or high virtualities). The scale $Q_s$ in Eqs.~(\ref{CA5},\ref{g-eff}), is generally called the saturation scale. For a more precise definition of the saturation scale, see Sec. III. In the CGC dipole model, the scale $Q_s$ is given by
\beq \label{qs}
Q_s\to Q_s(x)=\Lb \frac{x_0}{x}\Rb^{\frac{\lambda}{2}} \text{GeV}, \hspace{1cm} \text{CGC model.}
\end{equation}
The parameters $\mathcal{A}$ and
$\mathcal{B}$ in \eq{CA5} are determined uniquely from the matching of the dipole amplitude and
its logarithmic derivatives at $rQ_s=2$: 
\begin{equation}
\mathcal{A} =-\frac{N_0^2\gamma_s^2}{(1-N_0)^2\ln(1-N_0)}, \hspace{1cm} \mathcal{B} =\frac{1}{2}(1-N_0)^{-\frac{1-N_0}{N_0\gamma_s}}.  \label{AB}
\end{equation}
We recall that the BK equation predicts geometric scaling behavior \cite{gs}, namely that the amplitude $\mathcal{N}(x, r, b)$ is not a function of three independent variables, but only of one variable $\mathcal{Z}=rQ_s(x,b)$. For $r Q_s\,>\,2$ the dipole amplitude \eq{CA5} has this geometric scaling property \cite{gs},  while for $r Q_s<<2$, approaching the DGLAP regime, the diffusion term in the anomalous dimension $\gamma_{eff}$ in \eq{g-eff} violates geometric scaling, as it should.

In the CGC dipole model the amplitude does not depend on impact parameter. Therefore, the integral over impact-parameter in \eq{gp} is taken as an over-all normalization factor  $\sigma_0=2\int d^2b$, which can be obtained via a fit to data. The dipole cross-section can then be defined as $\sigma_{q\bar{q}}=\sigma_0 \mathcal{N} (x,r)$. In the CGC model, the parameter $\kappa=9.9$ is fixed at the LO BFKL value \cite{IIM}.  The central fits are obtained at a fixed $N_0=0.7$ \cite{watt-bcgc}, and then the other four parameters, namely $\gamma_s, x_0, \lambda, \sigma_0$, are obtained by a fit to the HERA data for the reduced cross-section, via a $\chi^2$ minimization procedure. 

Watt, Motyka and Kowalski \cite{watt-bcgc, ip-sat1} extended the CGC dipole model by introducing an impact-parameter dependence of the amplitude, the so-called b-CGC model. Here this dependence enters by introducing a $b$ dependence in the saturation scale\footnote{For simplicity we ignore the angle  or orientation dependence of the saturation scale, and assume that it depends only on the impact-parameter size, see Ref.\,\cite{or1}.} $Q_s$ in \eq{CA5},  
\beq \label{qs-b}
Q_s\to Q_{s}(x,b)\,\,=\,\,\Lb \frac{x_0}{x}\Rb^{\frac{\lambda}{2}}\,\exp\left\{- \frac{b^2}{4\gamma_s B_{CGC}}\right\} \text{GeV}, \hspace{1cm} \text{b-CGC model,}
\end{equation}
where now the parameter $B_{CGC}$, instead of $\sigma_0$ in the CGC dipole model, is a free parameter and is determined by other reactions, namely the $t$-distribution of the exclusive diffractive processes at HERA.  Following Ref.\,\cite{watt-bcgc}, in the b-CGC dipole model we let the parameter $N_0$ to be free along with $\gamma_s, x_0, \lambda$, and obtain their values via a fit to HERA data. In the b-CGC model, the parameters $\mathcal{A} $ and $\mathcal{B}$  are also given by \eq{AB}.

 One of the most salient features of the 
 CGC approach is the universality of particle production at small-x, so that all the complexity of the infinite-body problem at very high energy (or small x) is reduced to a one-scale problem, the hard saturation scale $Q_s$. This becomes the only dimensional relevant scale at which nonlinear gluon recombination effects start to become important. In this picture, it is quite natural to expect that the b-dependence of the scattering amplitude appears via the saturation scale, and that the saturation scale thereby becomes function not only of $x$, but also of the size of system or the impact-parameter profile of the system. The b-CGC dipole model keeps all the features of the CGC dipole model, including its geometric scaling property. 

The difference between the b-CGC and the IP-Sat models \cite{ip-sat1,ip-sat2} is illustrated in \fig{f-f}.  Although both models include saturation effects and depend on impact-parameter, the former is based on the non-linear BK equation, while the later is based on DGLAP evolution, incorporating the saturation effect via Glauber-Mueller approximation \cite{ip-sat1,ip-sat2}. Therefore, the underlying dynamics of two models are quite distinct.

\subsection{ Exclusive diffractive processes }
\begin{figure}[t]       
\includegraphics[width=0.48\textwidth,clip]{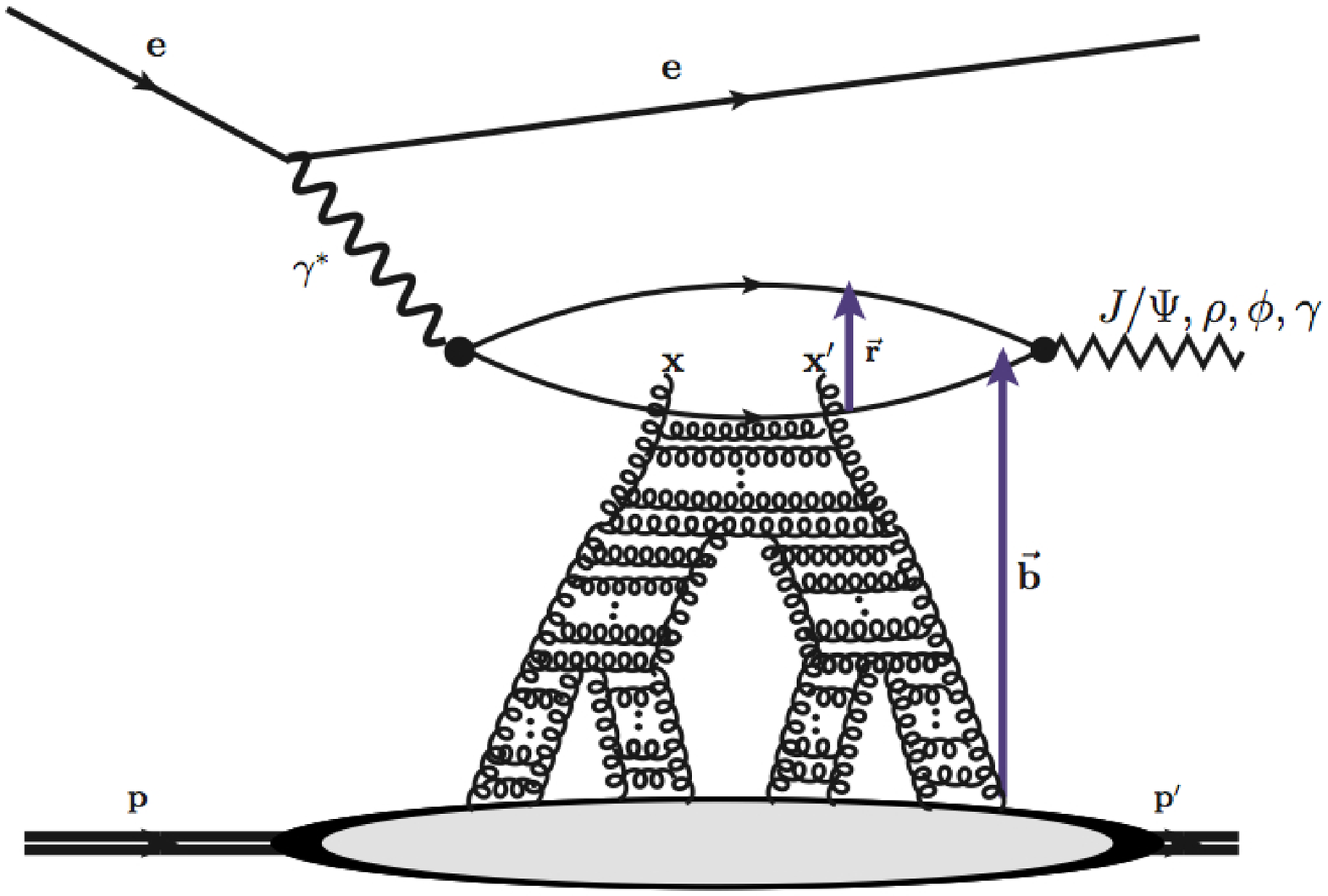}   
\includegraphics[width=0.48\textwidth,clip]{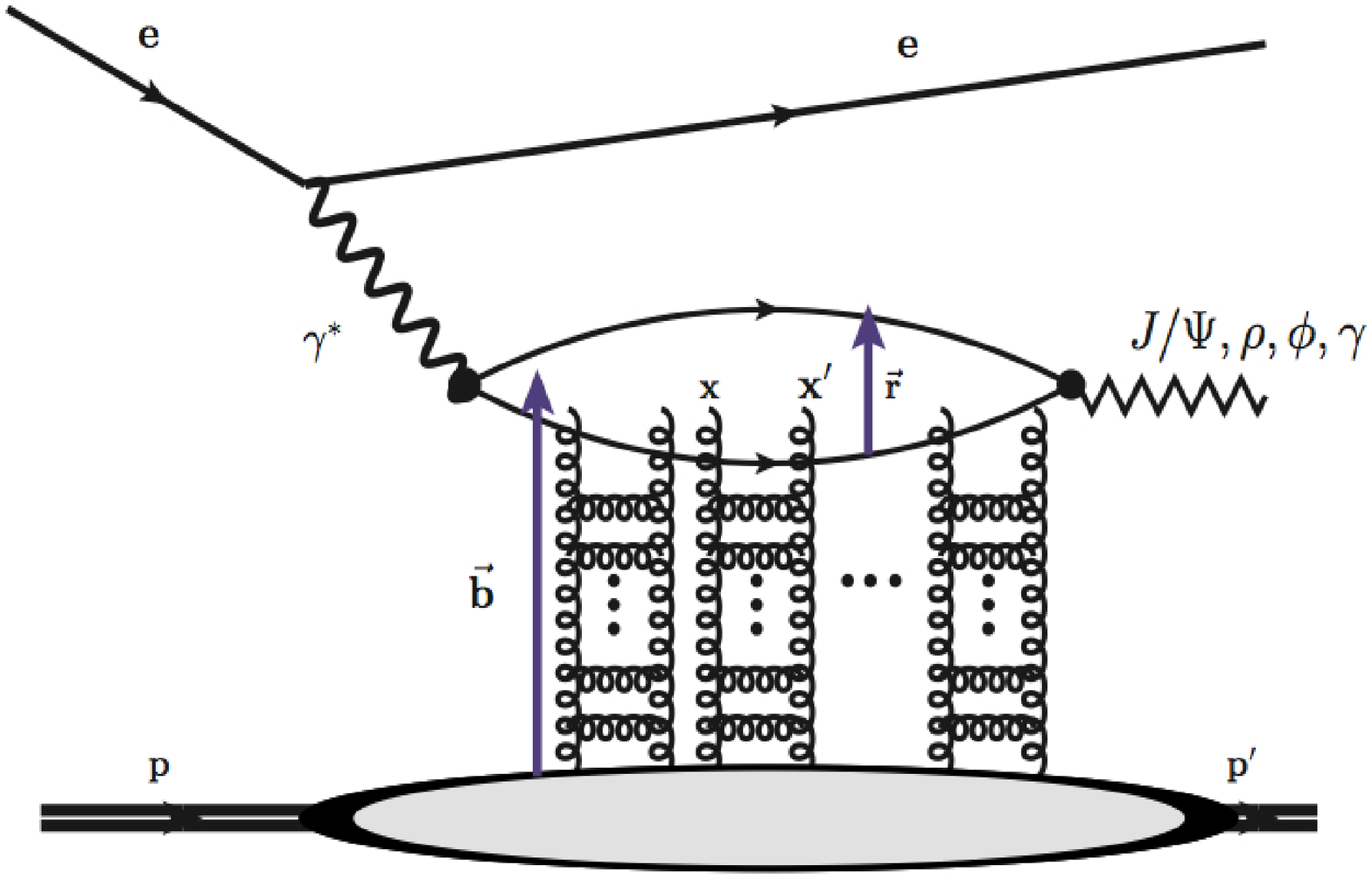}   
 \caption{The exclusive diffractive processes (with  $p\ne p^{\prime}$ or $t\ne 0$, and $x<<x^{\prime}<<1$)  in the b-CGC dipole model (left) and the IP-Sat dipole model (right) in the rest frame of the target.  }
\label{f-f}           
\end{figure}

The dipole formalism for the calculation of exclusive diffractive processes was thoroughly discussed in Refs.\,\cite{ip-sat1,ip-sat2} Here we only briefly discuss the main formulation.  Similar to the case of the inclusive DIS cross-section, the scattering amplitude for the exclusive diffractive process $\gamma^*+p\to E+p$, with a final-state vector meson $E=J/\Psi, \phi,\rho$  or a real photon $E=\gamma$  in DVCS, can be written in terms of a convolution of the dipole amplitude $\mathcal{N}$ and the overlap wave-functions of the photon and the exclusive final-state particle \cite{ip-sat1,ip-sat2}, see \fig{f-f}  

\begin{equation} \label{am-i}
  \mathcal{A}^{\gamma^* p\rightarrow Ep}_{T,L} = \mathrm{2i}\,\int\!\dif^2\vec{r}\int\!\dif^2\vec{b}\int_0^1\!\dif{z}\;(\Psi_{E}^{*}\Psi)_{T,L}(r,z,m_f,M_V;Q^2)\;\mathrm{e}^{-\mathrm{i}[\vec{b}-(1-z)\vec{r}]\cdot\vec{\Delta}}\mathcal{N}\left(x,r,b\right), 
\end{equation}
where $\vec{\Delta}^2=-t$, and $t$ is the squared momentum transfer. The phase factor $\exp\left(i(1-z)\vec{r}\cdot\vec{\Delta}\right)$ in the above equation is due to the non-forward wave-functions contribution \cite{bbb}.  Notice that in the overlap of wave functions $\Psi_{E}^{*}\Psi$ in  \eq{am-i}, summations over the quark helicities and over the quark flavor $f=u,d,s, c$ are implicit. The explicit form of the overlap wave functions $\Psi_{E}^{*}\Psi$ can be found in Refs.\,\cite{ip-sat1,ip-sat2}. One of the salient features of the dipole formalism is the fact that the color dipole is blind to the flavor, which allows to have a unified formalism for the description of both inclusive and diffractive vector meson production. In both cases, the underlying production mechanism is similar,  namely one needs to calculate the probability amplitudes of finding the color dipole of transverse size $r$ with impact parameter $b$ in the overlap wave function of photon (real or virtual)  and vector meson wave functions. 

There are several different prescriptions for modeling the vector meson wave functions. Following Refs.\,\cite{ip-sat1,ip-sat2,beta}, here we use the boosted Gaussian wave-functions \cite{beta} which were found to provide a very good description of exclusive diffractive HERA data \cite{watt-bcgc,ip-sat1,ip-sat2}. These wave-functions have no free parameters to be adjusted to the data that we want to describe here, since all its parameters have been already determined by a fit to other reactions, mainly to experimentally measured leptonic decay widths of vector mesons for the longitudinally polarized case \cite{ip-sat1}. In the case of the DVCS for real photon production only the transverse component of the overlap wave function contributes, and the wave function is generally better known compared to vector meson wave functions. 
For vector meson production, the dipole amplitude in \eq{am-i}  is evaluated at $x=x_{Bj}\left(1+M^2_V/Q^2\right)$, where $M_V$ denotes the mass of vector meson \cite{ip-sat1,ip-sat2}. 

The differential cross-section of the exclusive diffractive processes can then be written in terms of the scattering amplitude as \cite{ip-sat1,watt-bcgc}, 
\begin{equation}
  \frac{\dif\sigma^{\gamma^* p\rightarrow Ep}_{T,L}}{\dif t} = \frac{1}{16\pi}\left\lvert\mathcal{A}^{\gamma^* p\rightarrow Ep}_{T,L}\right\rvert^2\;(1+\beta^2) R_g^{2},
  \label{vm}
\end{equation}
where the factor $(1+\beta^2)$ takes into account the missing real part of amplitude (notice that the amplitude in \eq{am-i} is purely imaginary), with $\beta$ the ratio of the real to imaginary parts of the scattering amplitude, see e.g. \cite{ip-sat1,ip-sat2},
\begin{equation} \label{eq:beta}
  \beta = \tan\left(\frac{\pi\delta}{2}\right), 
\end{equation}
where we defined
\begin{equation} \label{delta}
 \delta \equiv \frac{\partial\ln\left(\mathcal{A}_{T,L}^{\gamma^* p\rightarrow Ep}\right)}{\partial\ln(1/x)}.
\end{equation}
The factor $R_g$ in \eq{vm} incorporates the skewness effect defined as $R_g=H(x,x)/H(2x,0)$, where $H$ is the off-forward gluon distribution \cite{ske}. 
 The skewness  factor takes into account the effect that the gluons attached to the $q\bar{q}$ can carry different light-front fractions $x,x^{\prime}$ of the proton \cite{ske,mrt}  (see \fig{f-f}).
At NLO level, in the limit that $x^{\prime}<<x<<1$ and  at small $t$,  assuming that the diagonal gluon density of target has the following generic power-law form
\begin{equation}
xg(x)=N_g x^{-\delta}, \label{g-a}
\end{equation}
the skewedness factor is given by \cite{ske}, 
\begin{equation} \label{eq:Rg}
  R_g(\delta) = \frac{2^{2\delta+3}}{\sqrt{\pi}}\frac{\Gamma(\delta+5/2)}{\Gamma(\delta+4)}.
\end{equation}
Notice that the assumed gluon-density profile (at small-x) given in \eq{g-a}, is generally consistent with the extracted gluon density in both the b-CGC and the IP-Sat dipole models. 

\begin{table}
  \centering
  \begin{tabular}{c|c|c|c|c|c}
    \hline\hline
      $Q^2/\text{GeV}^2$ bin &  $\gamma_s$  &$\sigma_0$/mb & $x_0$ & $\lambda$ & $\chisq$ \\ \hline
 $ [0.25,45]$ &  $ 0.762\pm 0.004$ & $21.85 \pm 0.03  $ & $6.226 \times 10^{-5}\pm 2.7\times 10^{-6} $&$0.2319\pm 0.0001 $&$351.3/297 =1.18$ \\ \hline
     $[0.75,650]$  & $ 0.719\pm 0.002$ & $24.064\pm 0.099$ & $ 2.22\times 10^{-5}\pm 1.95\times 10^{-8}$&$ 0.227\pm 5.8\times 10^{-5} $&$389.0/297 =1.3$ \\ \hline
  \end{tabular}
  \caption{Parameters of the CGC dipole model, determined from fits to data in the range $x\le 0.01$ and two bins of $Q^2\in [0.25, 45]\,\text{GeV}^2$ and $Q^2\in [0.75, 650]\,\text{GeV}^2$. The charm and light-quark masses are taken as $m_c=1.27$ GeV  and $m_u=10^{-2}\div 10^{-4}$ GeV, respectively (see the text for details).  }
  \label{t-1}
\end{table}

\begin{table}
  \centering
  \begin{tabular}{cc|c|c|c|c|c}
    \hline\hline
      $B_{CGC}/\mathrm{GeV}^{-2}$ &  $m_c$/GeV & $\gamma_s$  &$ N_0$ & $x_0$ & $\lambda$ & $\chisq$ \\ \hline
      5.5 & $1.27$ & $0.6599\pm  0.0003$ & $0.3358\pm 0.0004$ & $0.00105\pm 1.13\times 10^{-5}$&$0.2063\pm 0.0004 $&$368.4/297 =1.241$ \\ \hline
 5.5 &  $1.4$ & $ 0.6492\pm 0.0003$ & $ 0.3658\pm 0.0006$ & $0.00069 \pm 6.46\times 10^{-6}$&$0.2023\pm 0.0003 $&$370.9/297 =1.249$ \\ \hline
  \end{tabular}
  \caption{
Parameters of the b-CGC dipole model, determined from fits to data in the range $x\le 0.01$ and $Q^2\in [0.75, 650]\,\text{GeV}^2$. Results are shown for fixed light-quark masses $m_u=10^{-2}\div 10^{-4}$ GeV and two fixed values of the charm quark masses (see the text for details). }
  \label{t-2}
\end{table}

Unfortunately,  the calculation of the skewness factor $R_g$ deep inside the saturation region, where all twist contributions become important, is still an open problem \cite{ske}. Therefore, inevitably there will be  some possible mismatch between the approximation implemented in the calculation of the dipole amplitude and the skewness factor. This leads to an uncertainty with respect to the actual incorporation of the skewness correction at small $x$ in the dipole models\footnote{We checked that  the prescription given for the inclusion of the skewness effect in Eqs~(\ref{vm},\ref{delta}) is numerically very similar (for all observables) to the case in which we take $R_g^2\to R_g^{2\gamma_{\text{eff}}}$ and $\delta\to \lambda$. This indicates the robustness of our numerical results with respect to modeling the skewness effect, see also Ref.\,\cite{ip-conf}.}. Having this caveat in mind, the prescription given in \eq{vm} looks to be valid at all orders. Let`s for the sake of argument consider a simple case: at the leading-log approximation, for small dipole size $r$ in the color-transparency regime, the dipole amplitude is related to gluon structure function of the target, namely we have  $\mathcal{N}\approx r^2\alpha_s(r) xg(x,Q^2)$ \cite{cp}. Having this in mind, one can then readily extract the gluon density in the b-CGC model \eq{CA5} at the color-transparency regime $r Q_s\le 2$ with the anomalous dimension $\gamma_{\text{eff}}=1$, and we obtain $xg(x,Q^2)\approx Q_s^2\approx x^{-\lambda}$.  This has the same power-law like structure as in \eq{g-a}. Therefore, we can use the expression given in \eq{eq:Rg}  and consequently  take the value of $\delta$ in Eqs.(\ref{g-a},\ref{eq:Rg}) to be the same as the parameter $\lambda$ in the b-CGC dipole amplitude (obtained from a fit to HERA data) and the skewness contribution can be then included by producing the dipole amplitude at $r Q_s\le 2$ with the factor $R_g$.  One can generalize the above argument at the DGLAP  region with $\gamma_{\text{eff}}=1$ to the BFKL kinematics and beyond  by effectively extracting the value of $\delta$ via \eq{delta}, assuming again that the gluon density extracted from the dipole amplitude has the generic form given in \eq{g-a}.

Notice that the dipole amplitude is mainly determined from the reduced cross-section (or structure functions) alone; the choice of $R_g$ will only slightly affect the parametrization of the dipole via adjustment of the parameter $B_{CGC}$ in the impact-parameter profile of the saturation scale. We later quantify the uncertainty in extracting the parameter $B_{CGC}$ due to the inclusion of the skewness effect, and show that this uncertainty is very small.

\begin{figure}[t]       
\includegraphics[width=0.49\textwidth,clip]{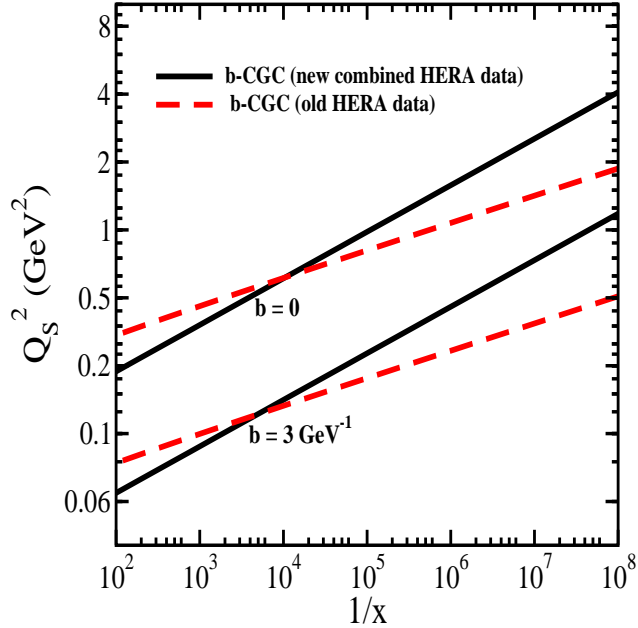}   
 \caption{The saturation scale extracted from the b-CGC model, with the parameter sets given in table \ref{t-2} (with $m_c=1.27$ GeV) and the old parameterization set from Ref.\,\cite{watt-bcgc}  as a function of $1/x$, at various impact-parameter $b$. }
\label{f-g1}           
\end{figure}     


\begin{figure}[t]       
\includegraphics[width=0.49\textwidth,clip]{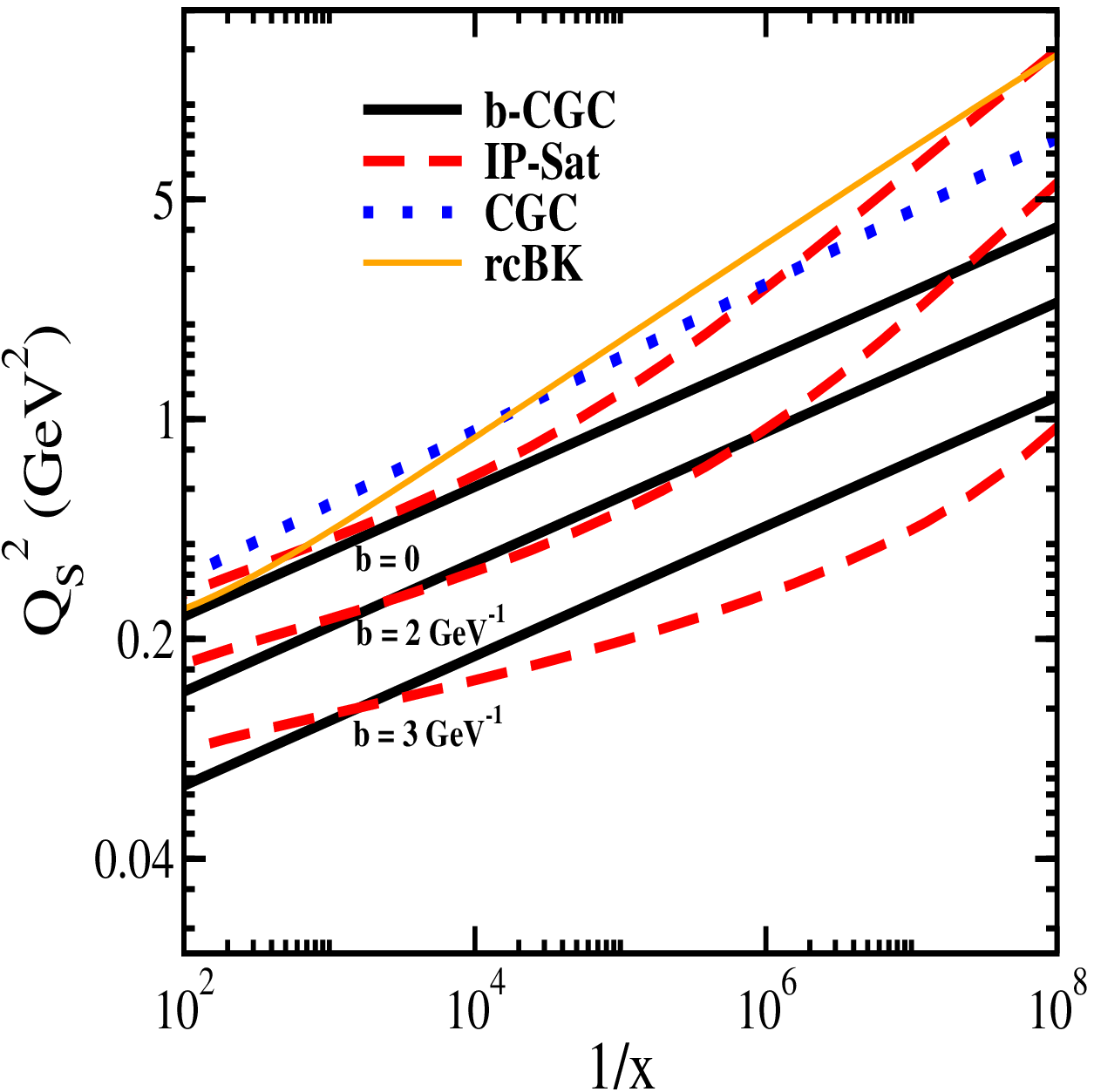}   
\includegraphics[width=0.49\textwidth,clip]{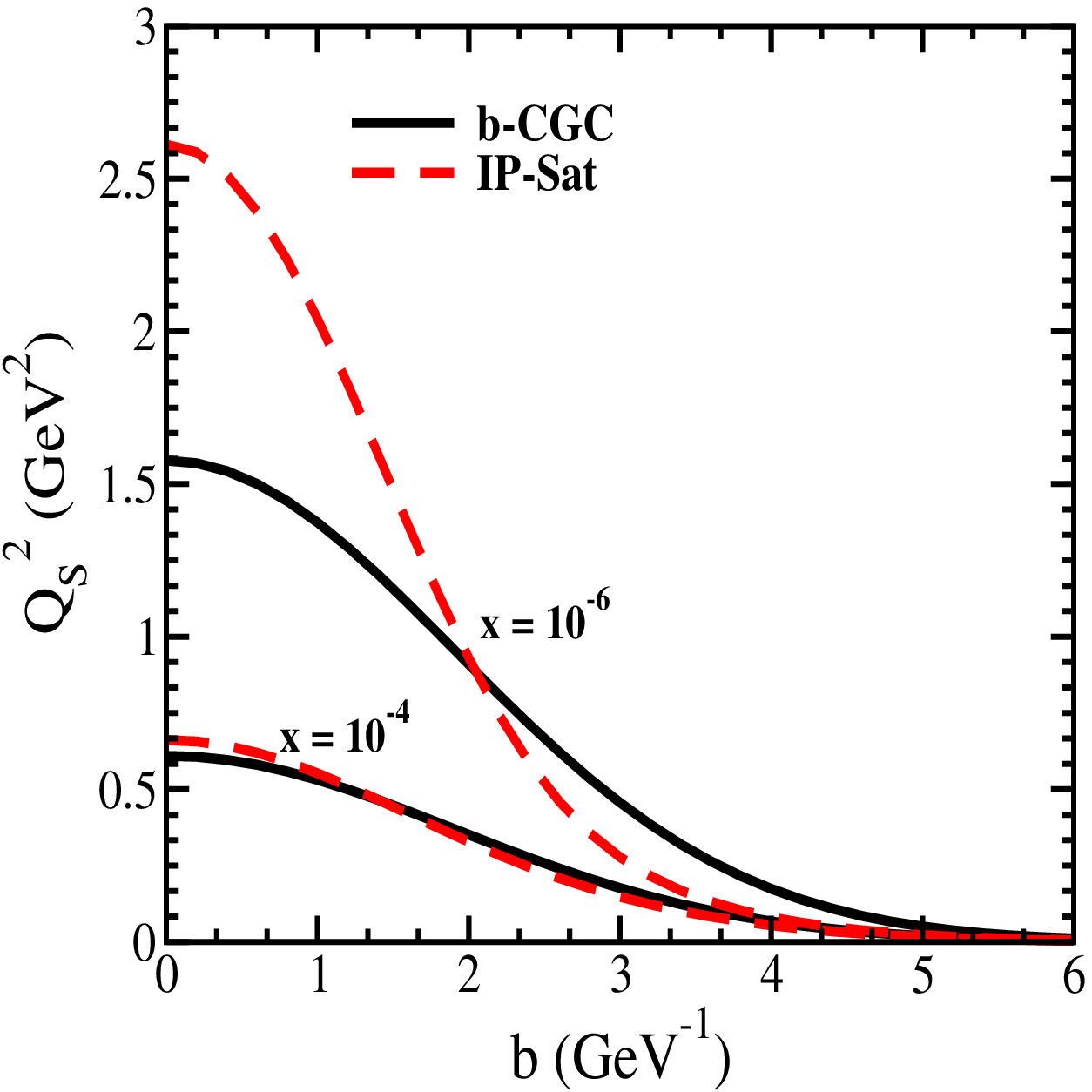}
 \caption{Left: The saturation scale in the b-CGC and the IP-Sat dipole models, as a function of $1/x$, at various impact-parameters $b$. For comparison we also show the impact-parameter independent saturation scale obtained from the CGC and the rcBK dipole model. For all models, the saturation scale was obtained from \eq{qs-d} with model parameters extracted from the combined HERA data.  Right: The saturation scale in the b-CGC and the IP-Sat models as function of the impact-parameter $b$, for various fixed values of $x$, obtained with the parameter set that includes charm mass $m_c=1.27$ GeV. }
\label{f-g2}           
\end{figure}     
\begin{figure}[t]       
\includegraphics[width=0.7\textwidth,clip]{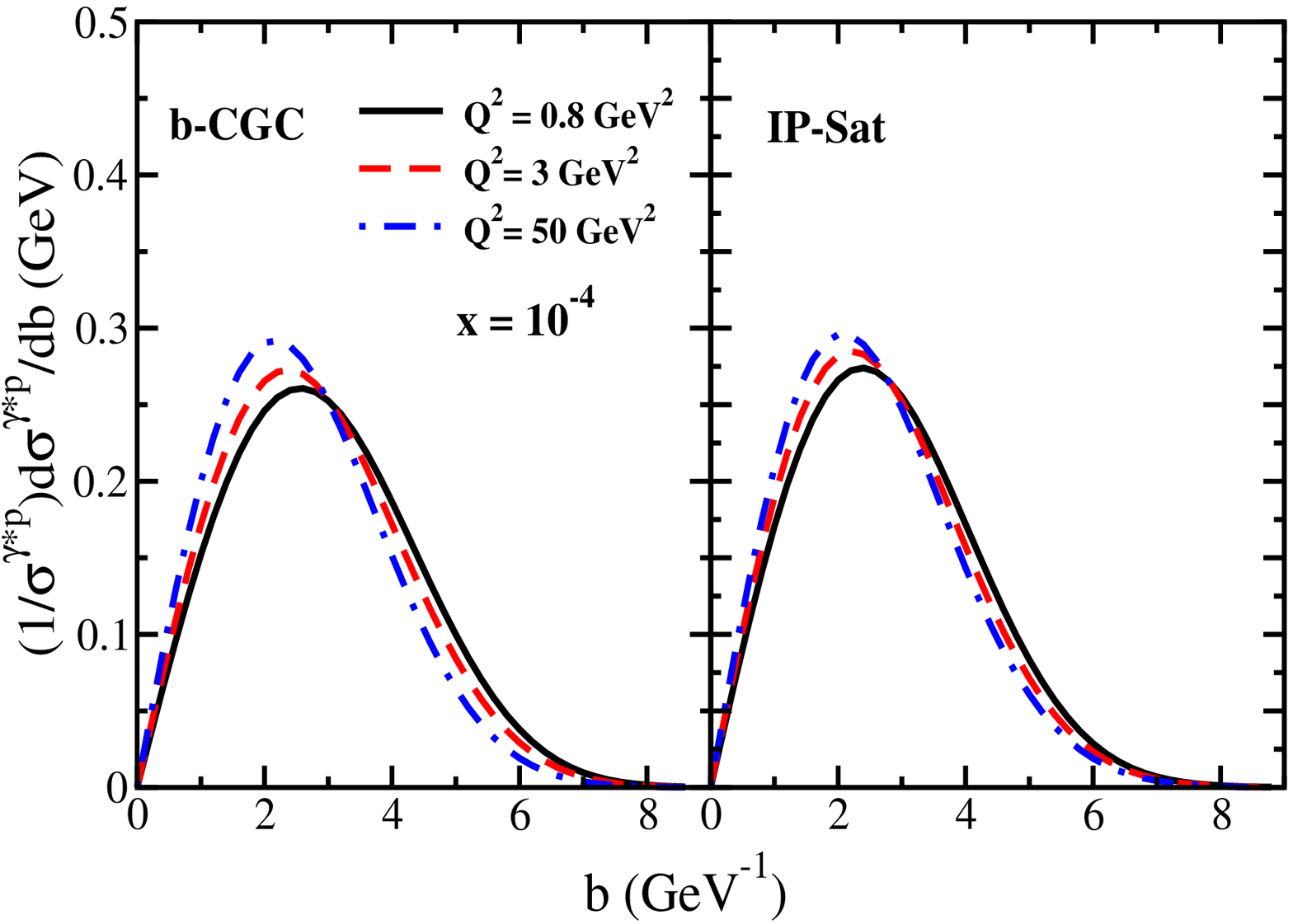}   
\includegraphics[width=0.7\textwidth,clip]{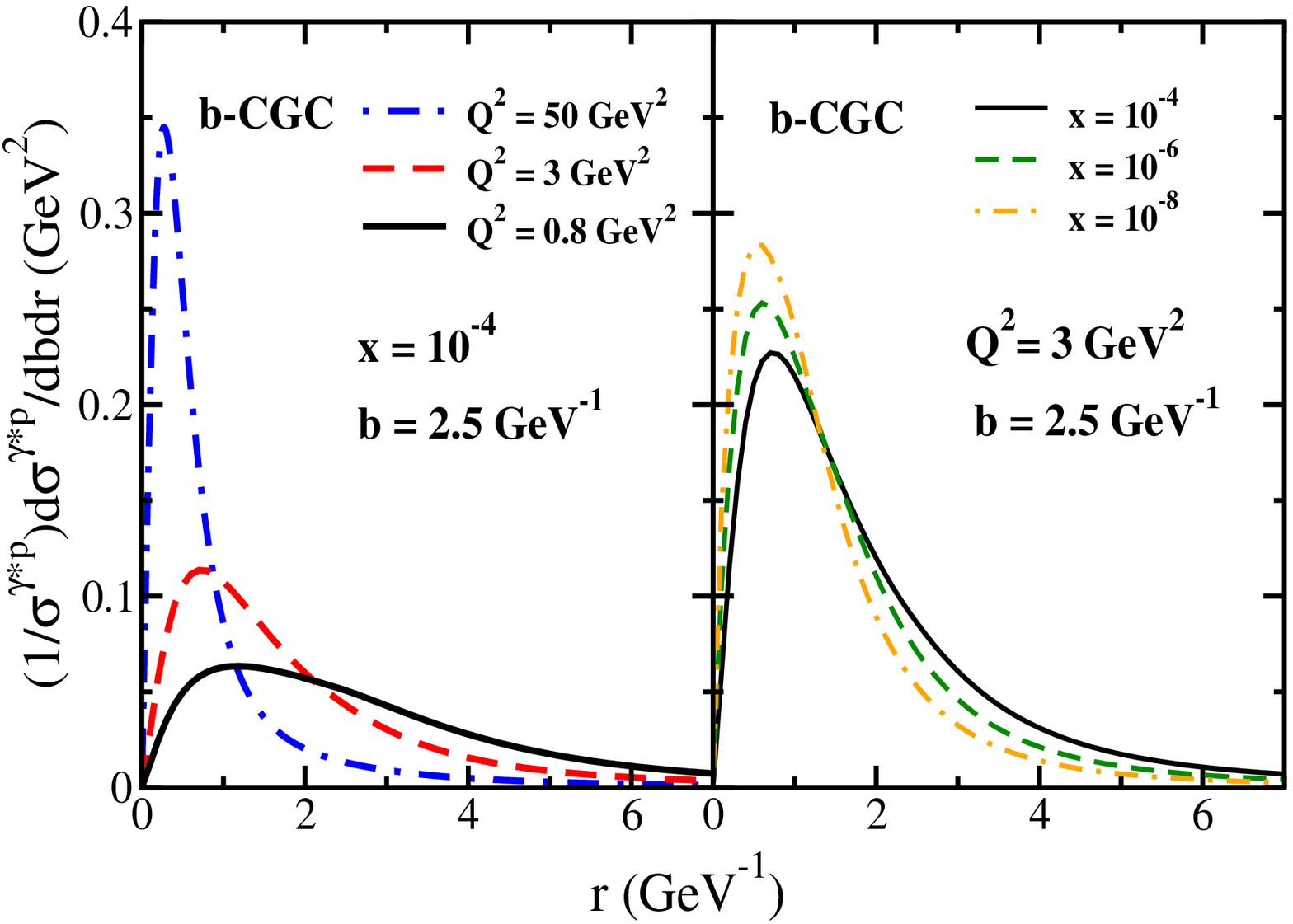}
 \caption{Top: The impact-parameter $b$ dependence of the total $\gamma^* p$ cross-section $\sigma^{\gamma^{*}p}$ , at  fixed $x$ and various $Q^2$, in the b-CGC and the IP-Sat dipole models. Lower: the dipole-size $r$ dependence of  the total $\gamma^* p$ cross-section, for fixed $x$ and $b$, but for various $Q^2$ (left); and  for a fixed $Q^2$ and $b$, but various $x$ (right), in the b-CGC dipole model. }
\label{f-g3}           
\end{figure}

\begin{figure}
  \includegraphics[width=0.8 \textwidth,clip]{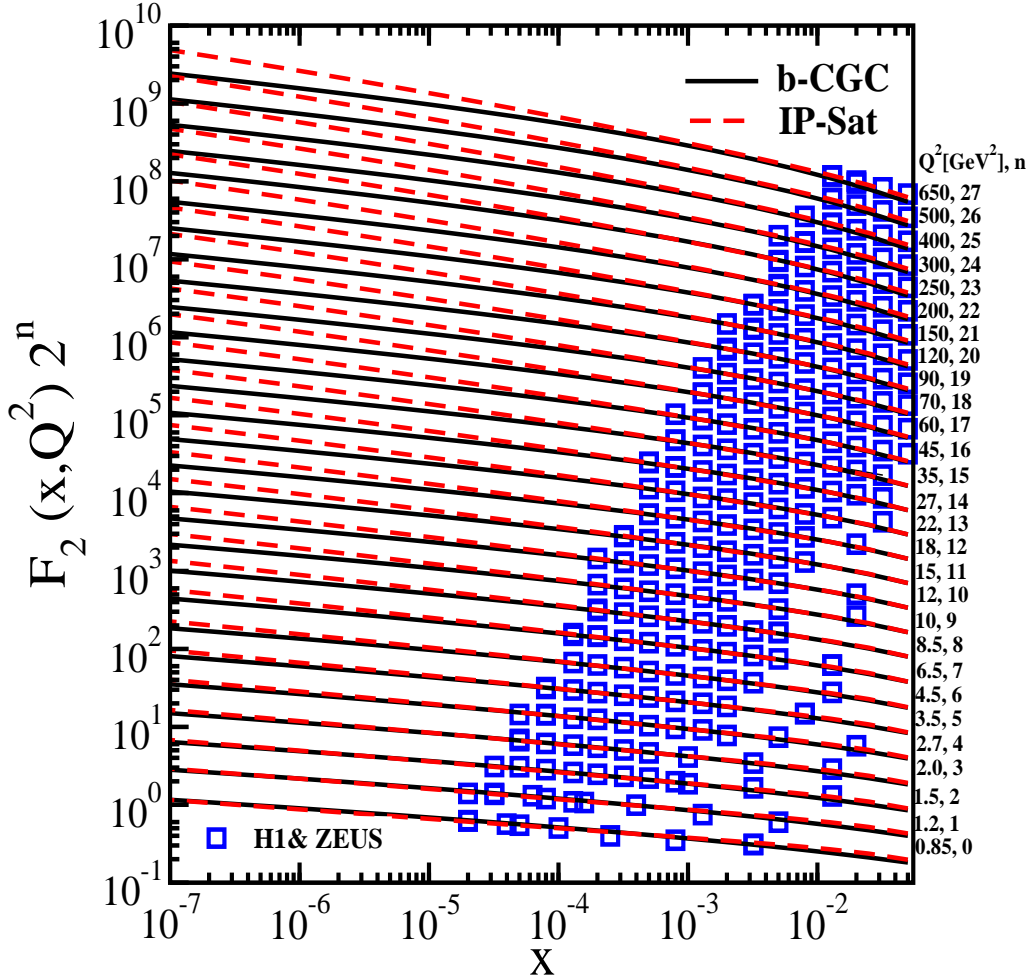}
  \caption{Results for the structure function $F_2(x,Q^2)$ as function of $x$, for various values of $Q^2$, in the b-CGC (solid line) and the IP-Sat (dashed line) dipole models. In order to separate data for each $Q^2$ from the others, the data and model results represented by the lines are multiplied by a factor $2^n$, with $n$ given on the right side of the plot. We used the parameter set of the b-CGC (in table \ref{t-2}) and the IP-Sat models with $m_c=1.27$ GeV. The experimental data are from H1 and ZEUS collaborations \cite{Aaron:2009aa}.  }
  \label{f-f2}
\end{figure}

\begin{figure}
\includegraphics[width=0.7\textwidth,clip]{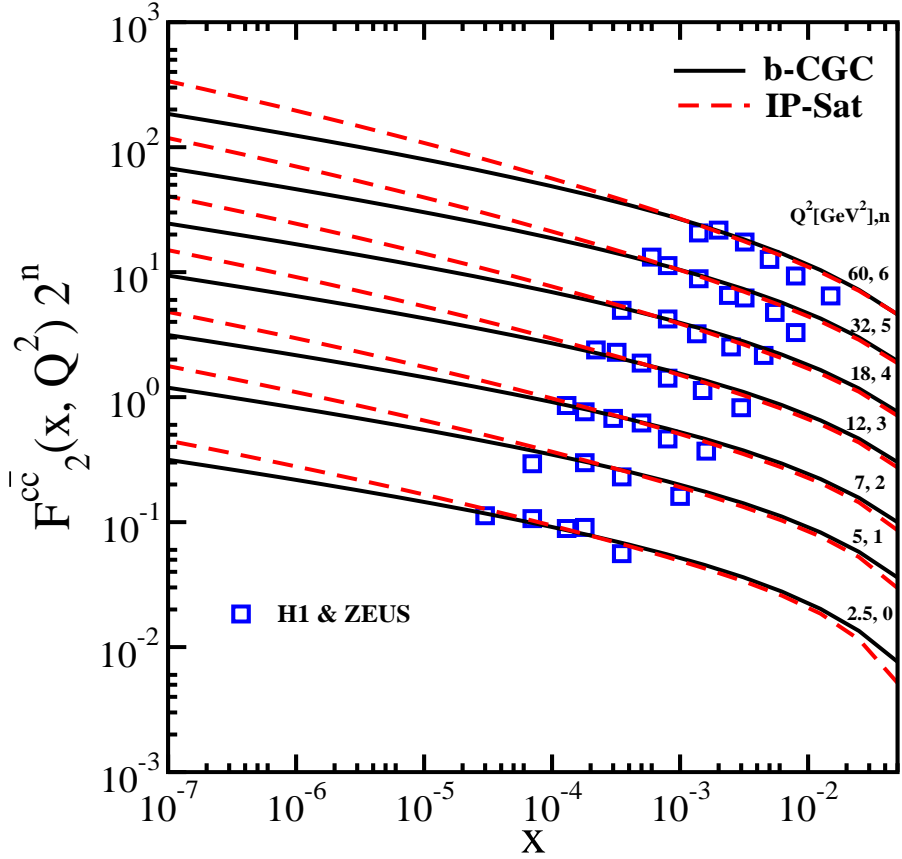}
  \caption{Results for the charm structure function $F_2^{c\bar{c}}(x,Q^2)$ as a function of $x$, for various values of $Q^2$, in the b-CGC (solid line) and the IP-Sat (dashed line) dipole models. We used the parameters set of the b-CGC (in table \ref{t-2}) and the IP-Sat models with $m_c=1.27$ GeV. The experimental data points are from the recently released combined  data sets of the H1 and ZEUS collaborations \cite{Abramowicz:1900rp}, assuming that $\sigma_r^{c\bar{c}}\approx F_2^{c\bar{c}}$ (see the text). }
  \label{f-f2c}
\end{figure}

\begin{figure}
  \centering
  \includegraphics[width=0.7\textwidth,clip]{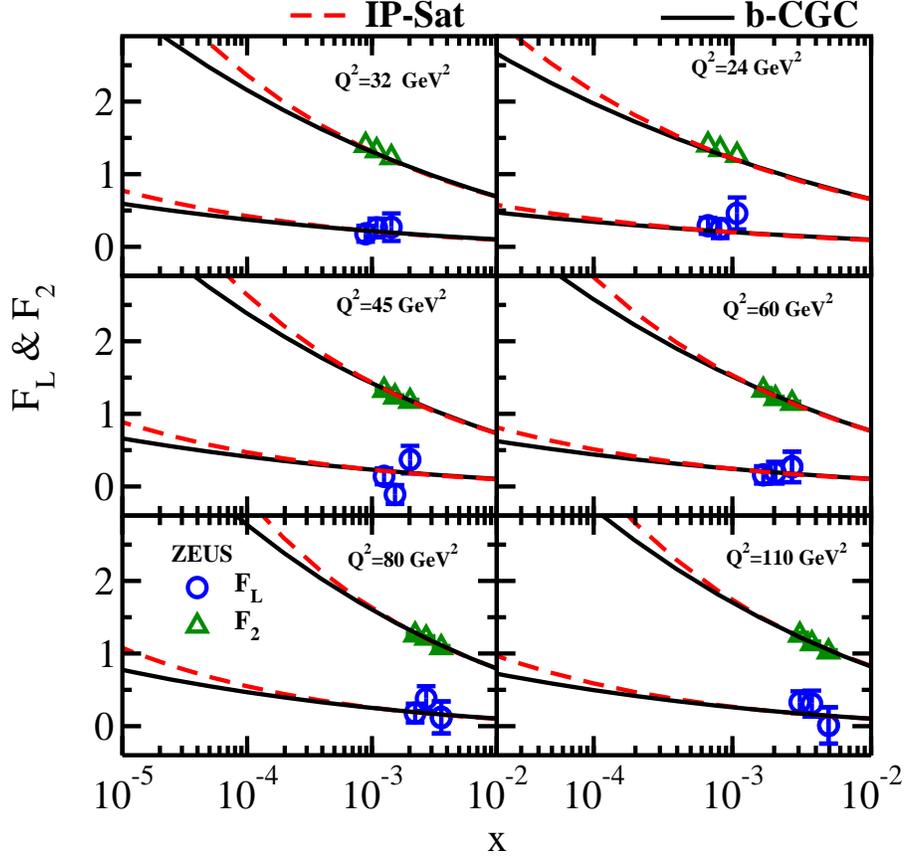}
  \caption{ Results for the longitudinal structure function $F_L(x,Q^2)$ and the structure function $F_2(x,Q^2)$, as  functions of $x$, for various values of $Q^2$, in the b-CGC (solid line) and the IP-Sat (dashed line) dipole models. We used the parameter set of the b-CGC (in table \ref{t-2}) and the IP-Sat models with $m_c=1.27$ GeV. The experimental data are from the ZEUS collaboration \cite{Chekanov:2009na}. }
  \label{f-fl}
\end{figure}

\begin{figure}
  \centering
  \includegraphics[width=0.6\textwidth,clip]{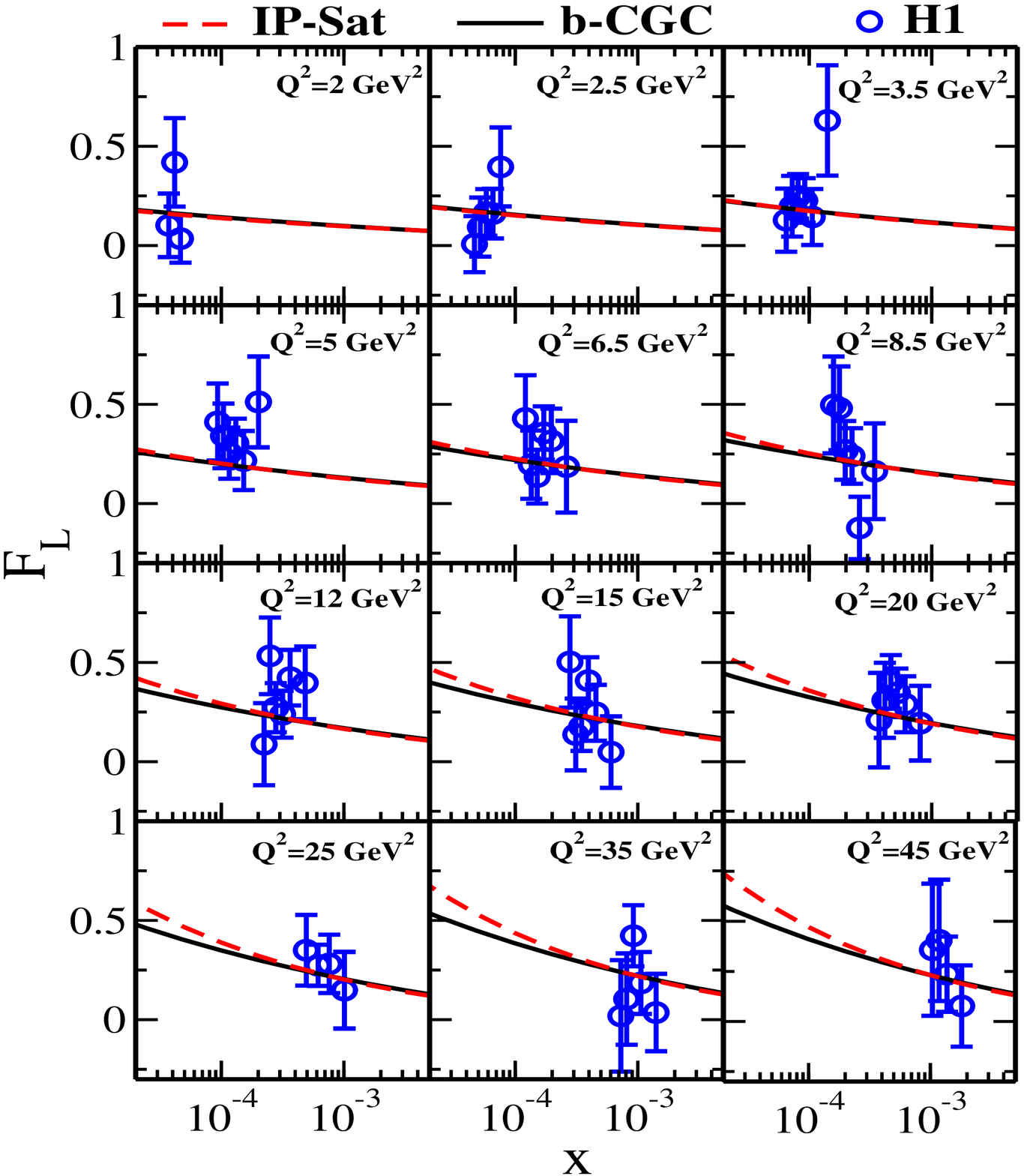}
  \caption{ Results for the longitudinal structure function $F_L(x,Q^2)$ as  functions of $x$, for various values of $Q^2$, in the b-CGC (solid line) and the IP-Sat (dashed line) dipole models. We used the parameter set of the b-CGC (in table \ref{t-2}) and the IP-Sat models with $m_c=1.27$ GeV. The experimental data are from the H1 collaboration \cite{new-fl-h1}. }
  \label{f-fl-h1}
\end{figure}

\section{Numerical results and discussion}
In earlier analyses for obtaining a fit for the b-CGC and the CGC dipole models \cite{watt-bcgc,IIM} only ZEUS $F_2$ data was included into the $\chi^2$  calculation, H1 data was not included in the fit in order to avoid introducing a possible normalization parameter between old H1 and ZEUS data. The new combined data from H1 and ZEUS collaborations \cite{Aaron:2009aa} are free from this problem. Moreover, the new data are extremely precise with error bars as small as $\sim1\%$. We include in our fit the recently released data for the reduced cross-section $\sigma_r$ from the combined analysis of the H1 and ZEUS collaborations \cite{Aaron:2009aa}. We also include data for structure functions from other experiments such as muon DIS data from E665 \cite{Adams:1996gu} and NMC~\cite{Arneodo:1996qe}. We note that the inclusion of those data to the $\chi^2$ minimization do not greatly change the quality of the fit and the values of free parameters. We should emphasize further that from HERA, we only include reduced inclusive DIS cross-section alone in our $\chi^2$ calculation. With the extracted parameters, we then confront model predictions for $F_2$, $F_L$ and $F_{2}^{c\bar{c}}$ with the HERA data (these data were not included into the fit).  In the calculation of $\chi^{2}$, the statistical and systematic experimental  uncertainties are added in quadrature.

In both the CGC and the b-CGC dipole model we have altogether, $4$ free parameters, which we fix via  a fit to experimental data by a $\chi^2$ calculation. Following Ref.\,\cite{watt-bcgc} , we do not allow the parameter $B_{CGC}$ in the b-CGC model to vary in addition to other parameters in the $\chi^2$  minimization algorithm, but we adjust it iteratively in order to obtain a good description of the $t$-dependence of the exclusive diffractive processes.

 The results of our fit in both the CGC and the b-CGC dipole models become stable with light quark mass ($m_u=m_d=m_s$) equal to current quark mass values of a few MeV, namely $m_u\approx 10^{-2}\div 10^{-4}$ GeV. This is due to the fact that here the saturation scale is larger than the confinement scale, and it thereby screens off the sensitivity to the infrared dynamics.  If following the old analysis \cite{watt-bcgc,IIM,soy}  we take $m_u=0.14$ GeV in the CGC model, the parameters of our fit change dramatically and the $\chi^2/d.o.f$ increases by about $8\%$.  Therefore, the new combined H1 and ZEUS data put a tougher constrain on the preferred value of quark mass in DIS, and it seems data prefer very light quark masses, in sharp contrast to the conclusion from the old data analysis. This feature was also recently observed in combined HERA data analysis in the IP-Sat dipole model \cite{ip-sat2}.

In table \ref{t-1} we show the results of our parameters fit in the CGC dipole model, assuming a fixed charm quark mass $m_c=1.27$ GeV. Following the previous analysis of Refs.\,\cite{watt-bcgc,IIM,soy}, in our $\chi^2$ calculation we first include data in the range $x\le 10^{-2}$ and $0.25\leq Q^2[\text{GeV}^2] \leq 45$,  and then we also consider data in $0.75\leq Q^2[\text{GeV}^2]\leq 650$ and  $x\le 10^{-2}$.  Altogether we include 263 experimental data points for $\sigma_r$ from the combined H1 and ZEUS data in neutral current unpolarized $e^{\pm}p$ scattering  \cite{Aaron:2009aa}, and 39 experimental data points from E665 \cite{Adams:1996gu} and NMC~\cite{Arneodo:1996qe}  structure function analysis. In the CGC model, the $\chi^2$ is rather stable with respect to the $Q^2$ lower cut in data bin selection (denoted by $Q^2_{\text{min}}$), while it worsens by enhancing the $Q^2$ upper cut. In the b-CGC model,  we have an opposite situation, namely changing the upper limit cut on $Q^2$ will not greatly influence  our fit, although increasing $Q^2_{\text{min}}>0.5\, \text{GeV}^2$ improves the fit and leads to a stable fit. Notice that our model, which is based on weak coupling dynamics, is unreliable at very low virtualities of the order or below the saturation scale. The saturation scale in both the CGC and the b-CGC dipole models, within the range of HERA kinematics, is rather small, about $0.2<Q_S[\text{GeV}]<1$ (see Figs.\,(\ref{f-g1},\ref{f-g2})). The CGC dipole model is more stable with respect to variations in the lower $Q^2_{\text{min}}$, due to the fact that the saturation scale in the CGC dipole model is generally larger than in the b-CGC model (even at b=0). However, the larger saturation scale in the CGC model, which is blind to impact-parameter, should be taken with care since the typical impact-parameter probed in the DIS is about $b\approx 1\div 4~\text{GeV}^{-1}$  (see \fig{f-g3} and related discussion below).  We found that, similar to what happened in the IP-Sat model, our fit results in the b-CGC model become stable with $Q^2_{\text{min}}\approx 0.75 \,\text{GeV}^2$. Therefore, in the b-CGC model, in the $\chi^2$ calculation  we consider data only in a range of  $x\le 10^{-2}$ and $0.75 \leq Q^2[\text{GeV}^2] \leq 650$.

In general the minimization is more sensitive to light quark masses than to the charm mass. This is because  the charm data does not influence greatly the parameters of the model,  even if we include $\sigma_r^{c\bar{c}}$ data in our $\chi^2$ calculation. In table \ref{t-2}, we show our fits for the b-CGC model, with two different charm masses $m_c=1.27$ and $1.4$ GeV. 

Using only the old HERA data one can not uniquely extract the value of the anomalous dimension $\gamma_s$ from a fit in the CGC dipole model. For example, it was shown in Ref.~\cite{watt-bcgc} that two parameter sets with two different value of $\gamma_s=0.61, 74$ give equally a good $\chi^2$ (see also \cite{soy}). The recent combined HERA data seem to favor a slightly higher value for $\gamma_s=0.76$, for the same $Q^2$ cut in data bin as in Refs.~\cite{watt-bcgc,IIM,soy}.  However, it is clear from tables \ref{t-1} and \ref{t-2} that the inclusion of impact-parameter dependence in the CGC model (namely the b-CGC model) reduces the extracted value of the anomalous dimension to $\gamma_s\approx 0.65$. This effect is also seen in the $\gamma_s$ value extraction from old HERA data in the b-CGC model, although a significantly lower value of about $\gamma_s=0.46$ was obtained \cite{watt-bcgc}.  We recall that the LO BFKL value of $\gamma_s$ is about  $0.63$ \cite{IIM}. However, one should bear in mind that in the b-CGC model, because of impact-parameter dependence, we intrinsically incorporate non-perturbative physics that is not present in the BFKL dynamics.

The saturation scale is a momentum at which the forward dipole-target scattering amplitude $\mathcal{N}$ rapidly raises with decreasing $x$ and the amplitude $\mathcal{N}$ becomes sizable, such that non-linear gluon recombination effects start to become as important as the gluon radiation.  
Following Refs.\,\cite{ip-sat1,ip-sat2,watt-bcgc}, we define the saturation scale $Q_S^2=2/r_S^2$, where $r_S$ is the saturation radius, as a scale where the dipole scattering amplitude has the value 
\beq
\mathcal{N}(x,r_S=\sqrt{2}/Q_S,b)=1-\exp(-1/2)=0.4. \label{qs-d}
\eeq
 Notice that in both the CGC and the b-CGC models, the saturation scale $Q_S$, defined via \eq{qs-d}, differs from the scale parameter $Q_s$ (with lower subscript $s$)  given in Eqs.~(\ref{qs},\ref{qs-b}), although they are closely related. It is important to note that the saturation scale does not have a unique definition, and in literature different definitions for extracting $Q_S$ can be found. Nevertheless, the definition in \eq{qs-d} gives a useful baseline to compare relative magnitude of saturation scale in different models. 

In \fig{f-g1}, we compare the saturation scales extracted from the b-CGC model using the old parameters of Ref.\,\cite{watt-bcgc} and with the new parameters obtained in this paper, as a function of $1/x$ at various impact parameters $b$. It is clear that the saturation scales extracted from the old and the new combined data from HERA are different, and this difference becomes more sizable at very small x. This is mainly due to the different power-law behavior of the saturation scale extracted from old and new combined HERA data via \eq{qs-d}, and it is directly related to the fact that the parameter $\lambda$ is larger in the new fit ($\lambda=0.206$, see table \ref{t-2}) compared to the earlier analysis (with $\lambda=0.119$).  Moreover,  the parameter $B_{CGC}$ extracted from the t-slope of the diffractive processes is here smaller than the value obtained in Ref.\,\cite{watt-bcgc}. Therefore, the earlier b-CGC fits should be superseded by the fit given in table \ref{t-2}. 

 In \fig{f-g2} (left), we show the saturation scale as a function of impact parameter $b$, for different values of $x$ in the b-CGC and the IP-Sat models. We see that the saturation scale as a function of  $1/x$ grows relatively  faster for more central collisions ($b\approx 0$). Moreover, the saturation scale at different impact parameters can be significantly different, even by one order of magnitude. This non-trivial behavior shows the importance of the impact-parameter dependence of the saturation scale. It is remarkable that although the b-CGC and the IP-Sat models are different, both give similar saturation scales within the x-region that they have been fitted to the HERA data, namely within $x\in [10^{-2},10^{-5}]$. However, at smaller $x$ about $x<10^{-5}$, they become significantly different. This is because the power-law behavior of the saturation scale in the b-CGC and the IP-Sat models,  are different at different impact parameter values, and the growth of saturation scale with $1/x$ at more central impact-parameter is faster in the IP-Sat model. In  \fig{f-g2} (left) we also compare with the saturation scale extracted from impact-parameter independent models, such as the CGC model (with parameter given in table \ref{t-1}, first row) and the rcBK model with parameter set corresponding to $\gamma=1.119$ in Ref.\,\cite{rcbk}. Notice that all the curves in \fig{f-g2} were obtained via definition of the saturation scale given in \eq{qs-d}. 
The CGC and the rcBK dipole model generally give a larger saturation scale than the b-dependent dipole models, and this difference can be as large as one order of magnitude at peripheral impact-parameter. In \fig{f-g2} (right), we compare saturation scales extracted in both the b-CGC and the IP-Sat model, as a function of impact-parameter $b$, for different fixed values of x.  We see again that in the range of x in which these models give equal description of existing HERA data, the saturation scales in both models are very similar.  However, at smaller x, in the region that  we do not have data yet, these models give very different $Q_S$. We will show in the following that this leads to sizable different predictions for various observables in the b-CGC and the IP-Sat models at very small-x.   

In \fig{f-g2} top panel, we show the impact-parameter $b$ dependence of the total $\gamma^{*}p$ cross-section calculated by the b-CGC and the IP-Sat dipole models, at fixed x and various $Q^2$.  We see that the main contribution of the integrand in the structure functions and the reduced cross-section  at various virtualities $Q^2$ comes from $1\leq b [\text{GeV}^{-1}]\leq 4$, with a median value of about $2\div 3~\text{GeV}^{-1}$. Although the $b$ dependence of the dipole amplitude is different in the b-CGC and the IP-Sat models, remarkably both lead to the same conclusion that the typical $b$ probed in the total $\gamma^{*}p$ (and the structure functions) is about  $2\div 3~\text{GeV}^{-1}$. This feature is the same in both old and new combined HERA data \cite{watt-bcgc}, which clearly indicates that the dipole models without impact-parameter dependence overestimate the importance of the saturation effects in the HERA data and beyond.  

\fig{f-g2} lower panel shows the total $\gamma^{*}p$ cross-section as a function of dipole transverse size $r$, at various $Q^2$, but at a fixed value of x and b, in the b-CGC model. We see that, as expected, the typical relevant dipole size $r$ in the interaction depends on the virtuality $Q^2$, and at a larger $Q^2$ the main contribution of the total $\gamma^{*}p$ cross-section comes from dipoles with smaller transverse sizes $r$, while varying $x$ but keeping  $Q^2$ and $b$ fixed seems to have less impact on the typical dipole size $r$ probed in DIS.


With the parameters given in table \ref{t-2}, obtained from the $\chi^2$ calculation, we now compute the structure functions $F_2(x,Q^2)$, the charm structure function $F_2^{c\bar{c}}(x,Q^2)$ and  the longitudinal structure function $F_L(x,Q^2)$  in the b-CGC model, using Eqs.\,(\ref{f2},\ref{FL}), and compare to the combined HERA data. As emphasized previously, experimental data for $F_2$, $F_L$ and $F_2^{c\bar{c}}$  were not included in our fit and therefore this can be considered as a non-trivial consistency check of the model. We see in Figs.\,\ref{f-f2},\ref{f-f2c},\ref{f-fl} that the b-CGC model results are in good agreement with structure functions data, for a wide range of kinematics: for $Q^2\in [0.75, 650]\,\text{GeV}^2$ and $x\le 0.01$. For $F_2^{c\bar{c}}$ we confront the model results with new H1+ZEUS combined data  \cite{Abramowicz:1900rp}, assuming that $\sigma_r^{c\bar{c}}\approx F_2^{c\bar{c}}$. Notice that the contribution of $F_L^{c\bar{c}}$ to the reduced cross-section, originating from the exchange of longitudinally polarized photons, can be around or less than a few per cent, which is ignored in the kinematic range considered in \fig{f-f2c}. Unfortunately, the combined HERA data for  $F_L(x,Q^2)$ are not yet available. In  Figs.\,\ref{f-fl},\,\ref{f-fl-h1} we confront our model predictions  for  $F_L(x,Q^2)$ with existing ZEUS \cite{Chekanov:2009na} and H1 \cite{new-fl-h1} data. The ZEUS \cite{Chekanov:2009na} and H1 \cite{new-fl-h1,Aaron:2008ad} data for $F_L$ are consistent within the rather large error bars. In Figs.\,\ref{f-f2},\ref{f-f2c},\ref{f-fl},\ref{f-fl-h1} we also compare the structure functions results obtained from the b-CGC and the IP-Sat dipole models, and extend our model results beyond the kinematics of existing data, as predictions for future DIS experiments.  It is seen that both dipole models provide equally excellent descriptions of the existing data, while at very small $x$ and high virtuality $Q^2$, beyond the current DIS experimental kinematics, they are systematically different. This difference can be traced back to the different power-law behavior of the saturation scale in these models (see \fig{f-g2}). 

\begin{figure}[t]       
\includegraphics[width=0.49\textwidth,clip]{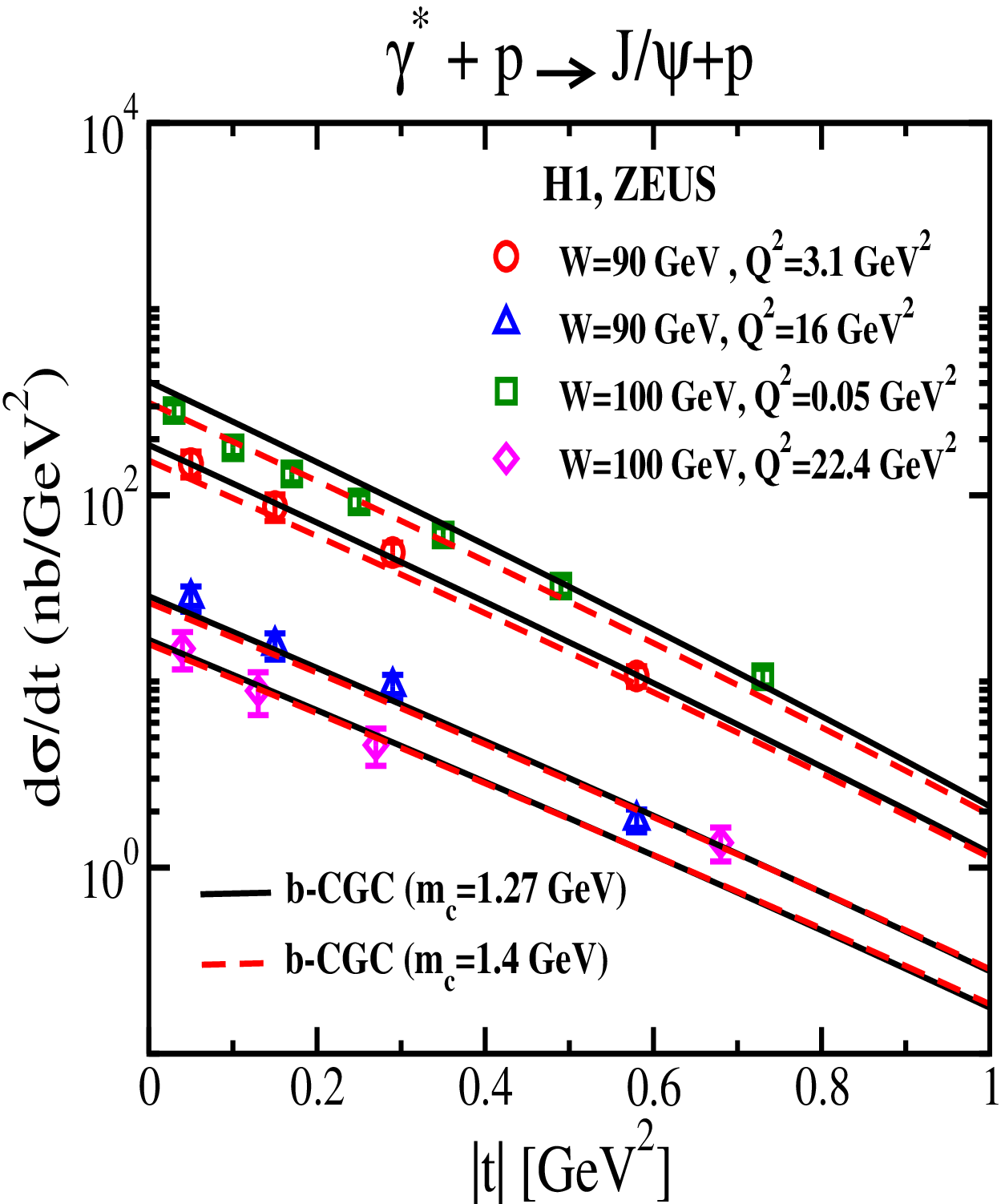}   
\includegraphics[width=0.49\textwidth,clip]{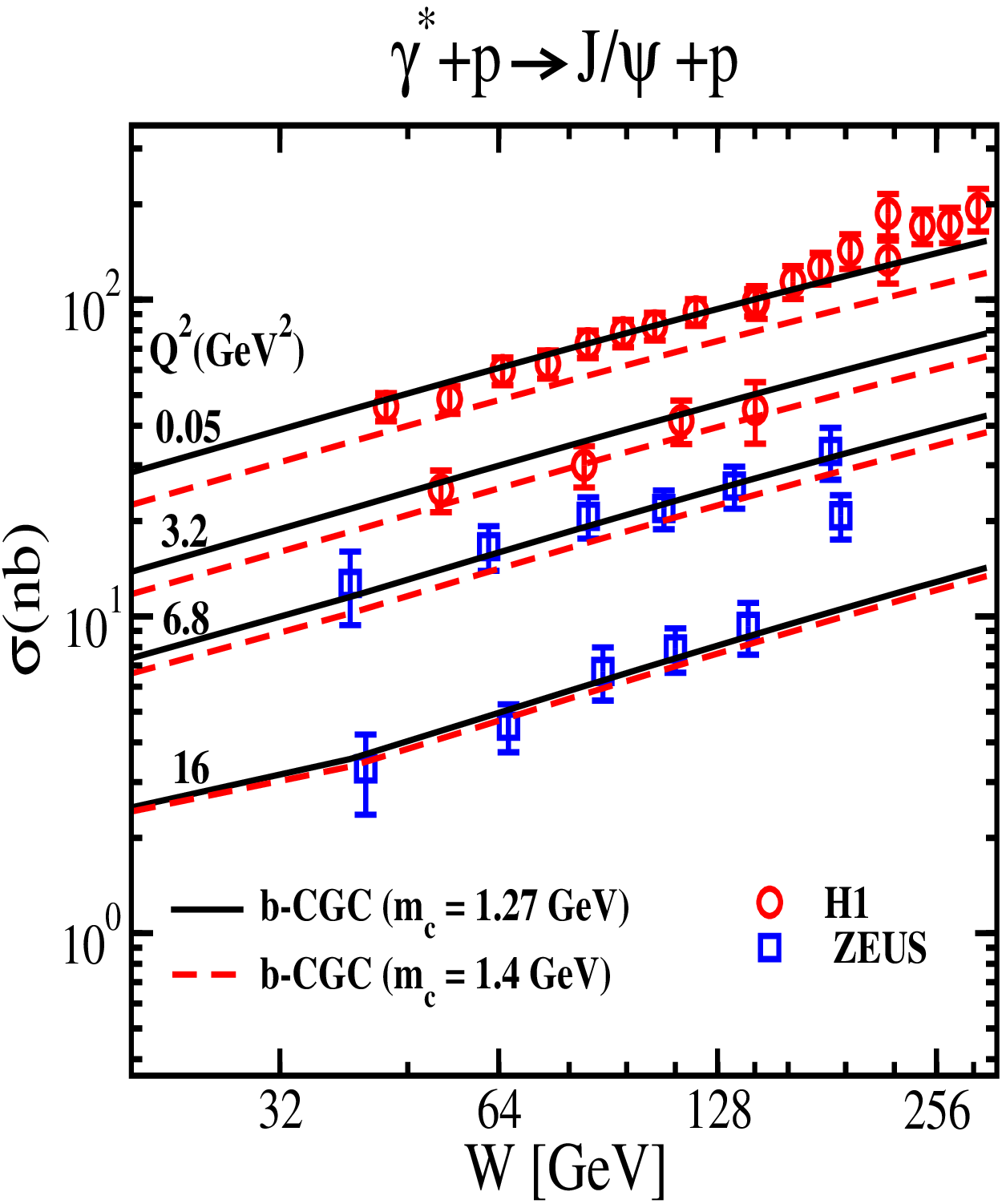}   
 \caption{Differential vector meson cross-sections for $J/\Psi$, as a function of $|t|$ (left) and $W$ (right). Data for various $Q^2$ are compared to results from the b-CGC model, using the two parameter sets given in table \ref{t-1}, with $m_c=1.27$ GeV (solid lines) and $m_c=1.4$ GeV (dashed lines).  The data are from the H1 and ZEUS collaborations \cite{Chekanov:2002xi,Chekanov:2004mw,Aktas:2005xu}.}
\label{f-ch}           
\end{figure}

\begin{figure}[t]       
\includegraphics[width=0.4\textwidth,clip]{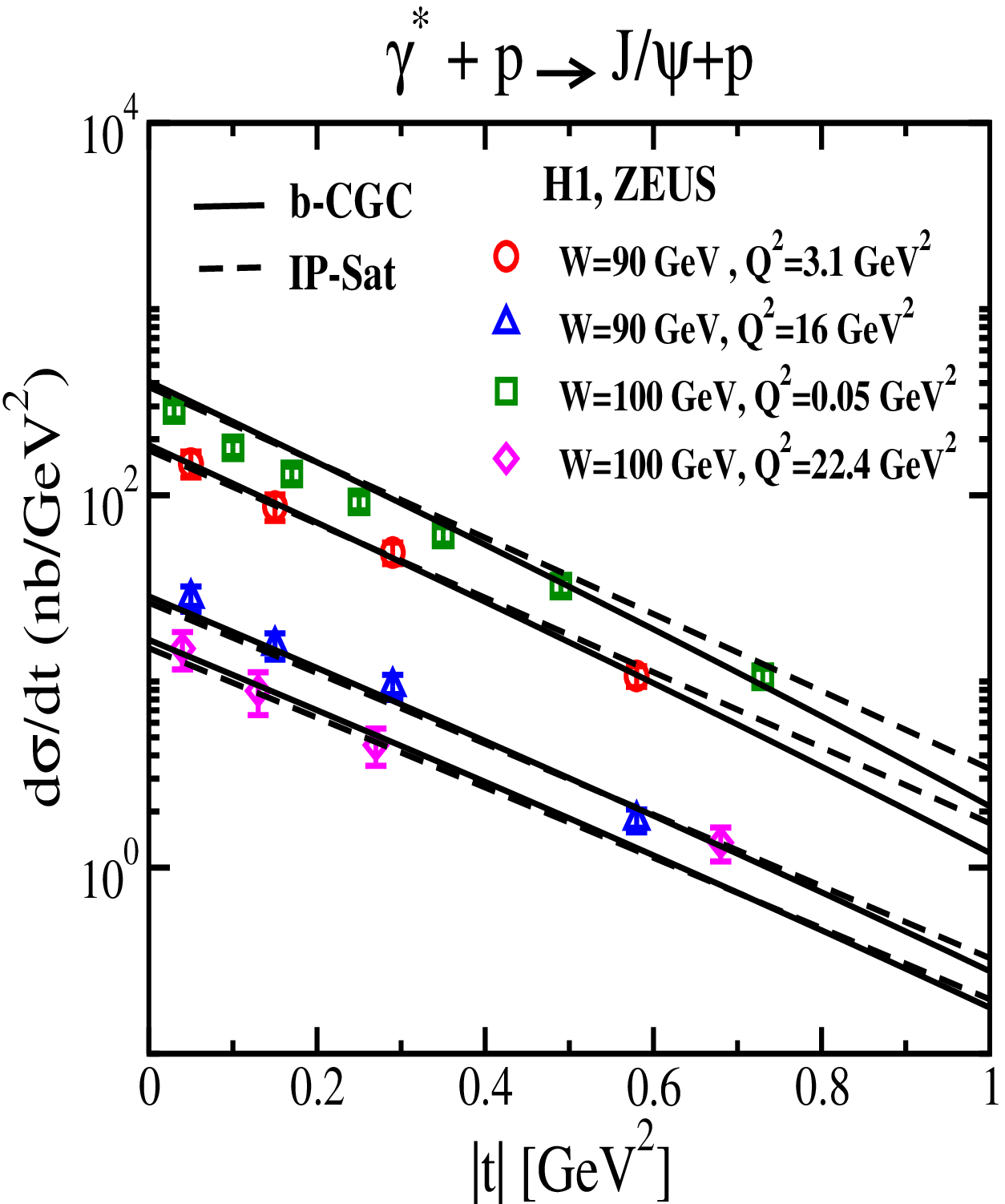}   
\includegraphics[width=0.4\textwidth,clip]{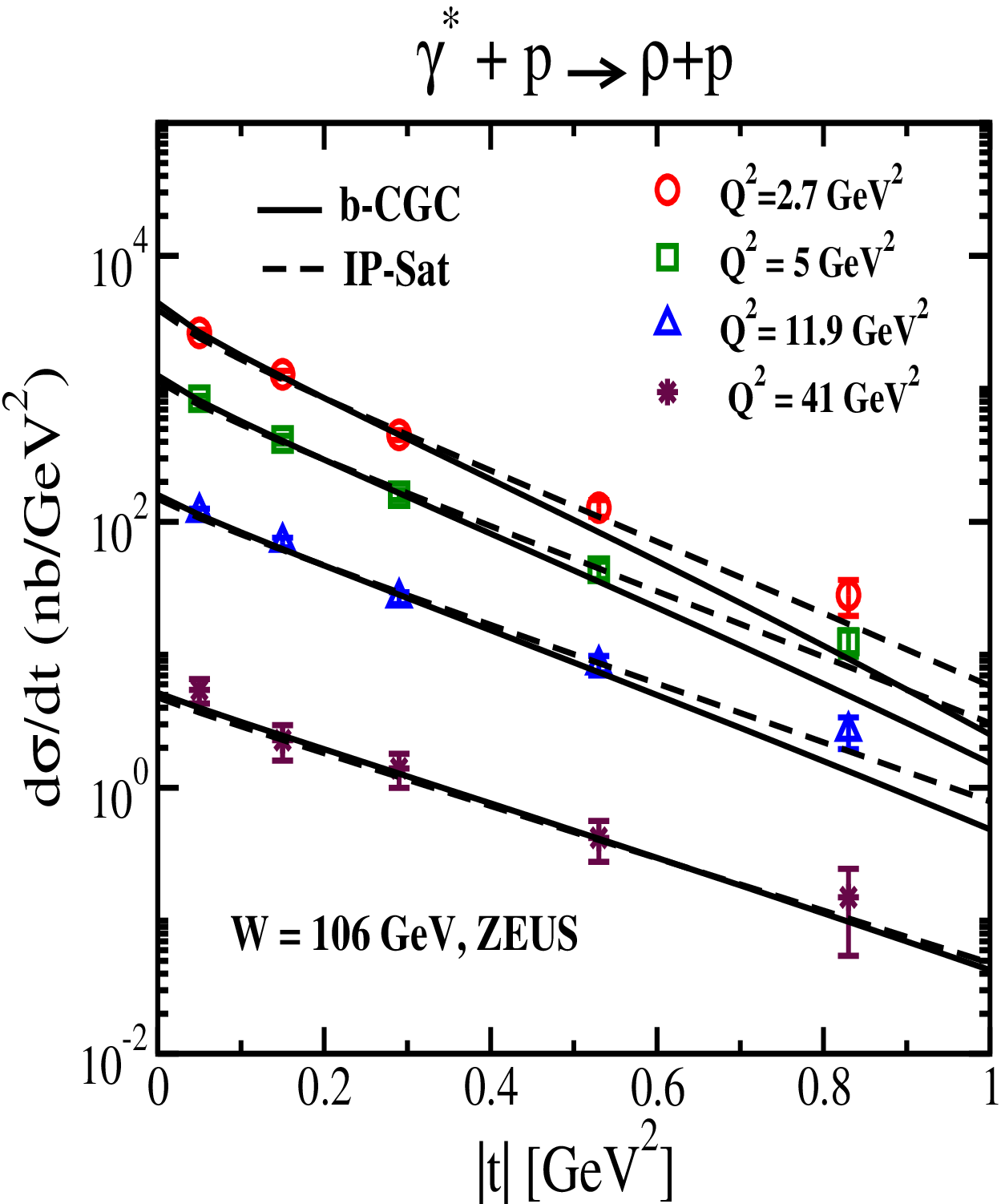}
\includegraphics[width=0.4\textwidth,clip]{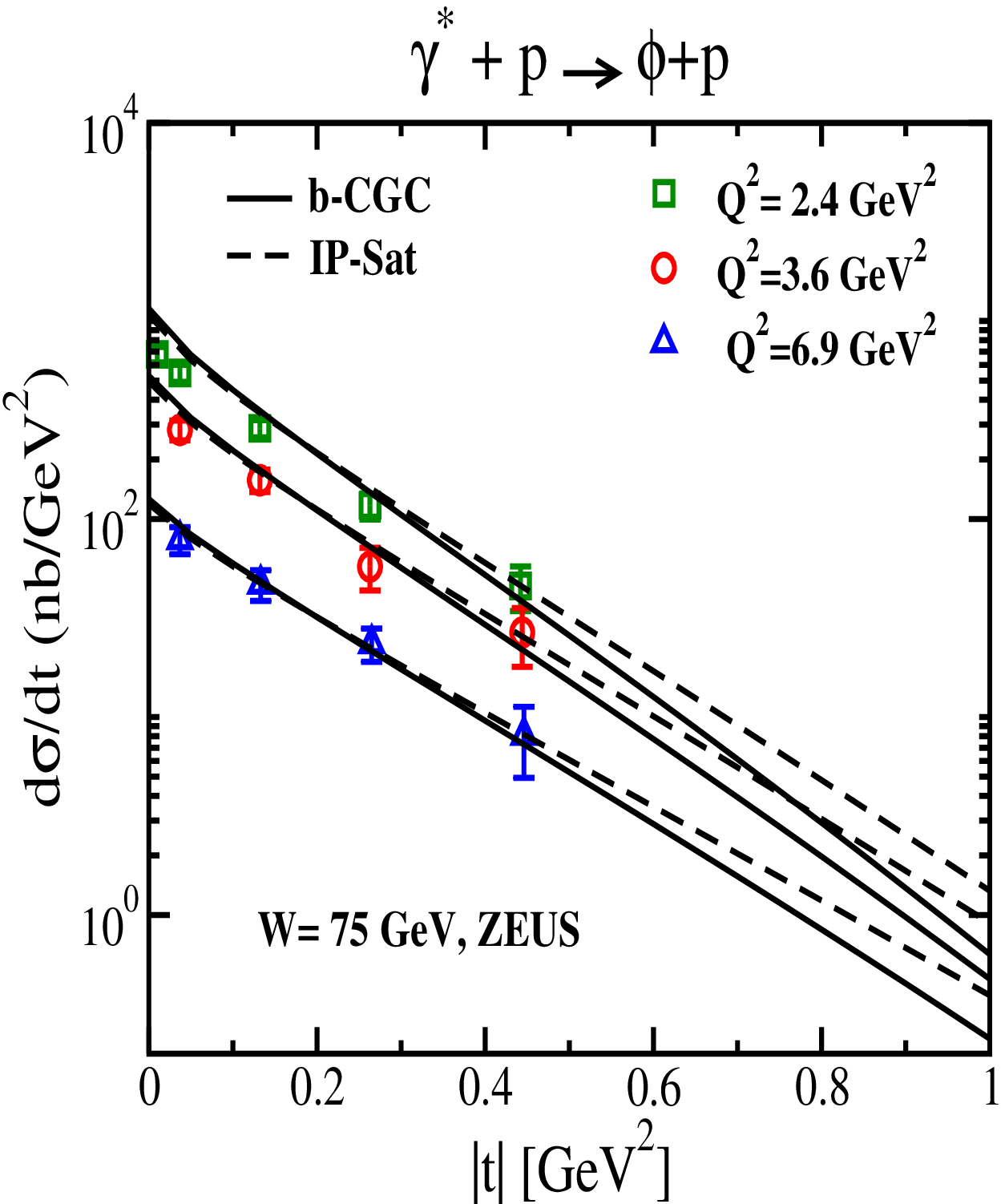}
\includegraphics[width=0.4\textwidth,clip]{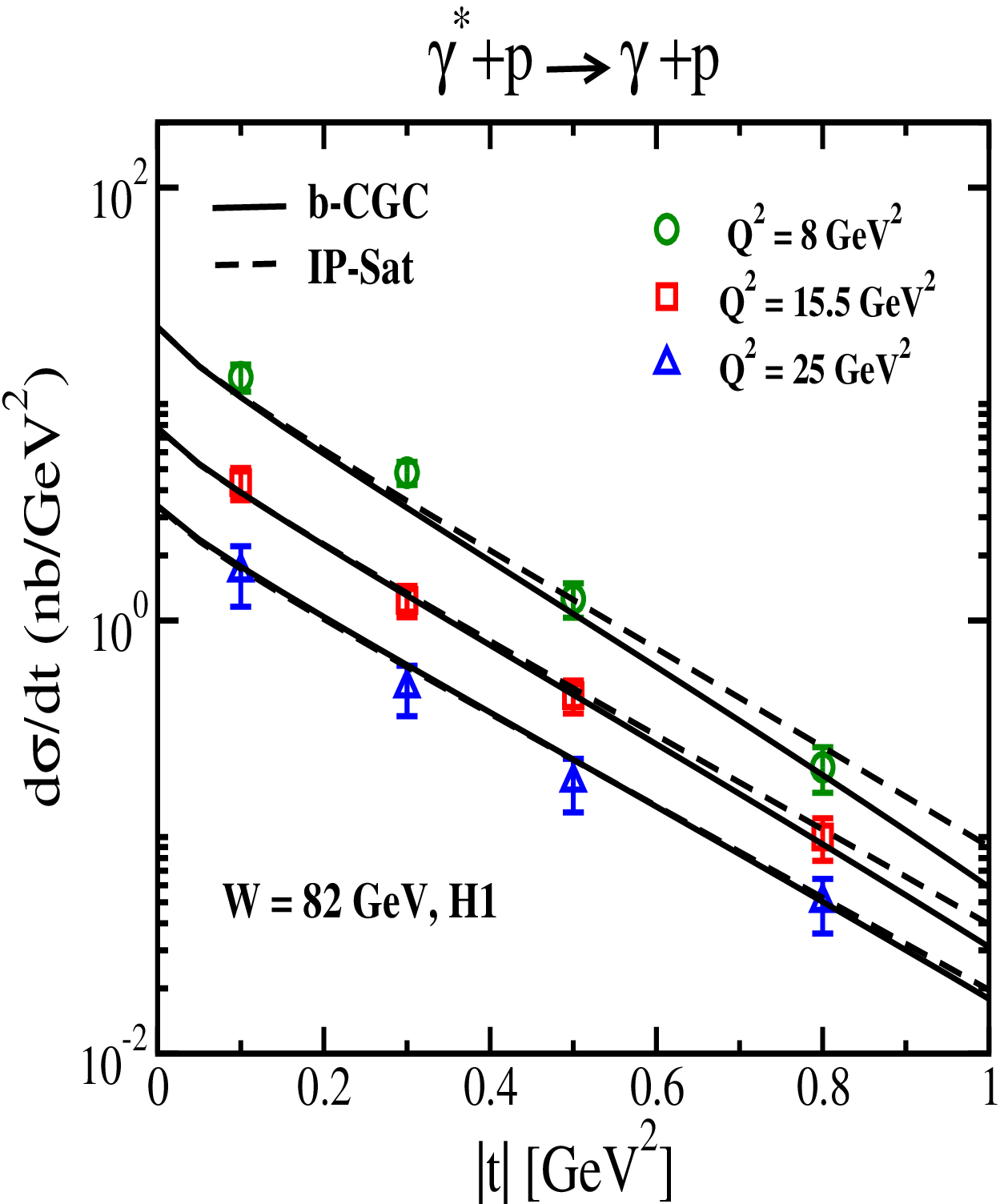}
 \caption{Differential vector meson cross-sections for $J/\Psi$, $\rho$, $\phi$ and DVCS, as a function of $|t|$. Data for a given $W$ with varying $Q^2$, are compared to the results from the b-CGC (solid lines) and IP-Sat (dashed lines) models, using the parameter sets with $m_c=1.27$ GeV in both models.  The data are from the H1 and ZEUS collaborations \cite{Chekanov:2002xi,Chekanov:2004mw,Aktas:2005xu,Chekanov:2005cqa,Aaron:2009xp,Chekanov:2007zr,Aaron:2009ac,Chekanov:2008vy}.  }
  \label{f-vt}
\end{figure}

\begin{figure}[t]       
\includegraphics[width=0.4\textwidth,clip]{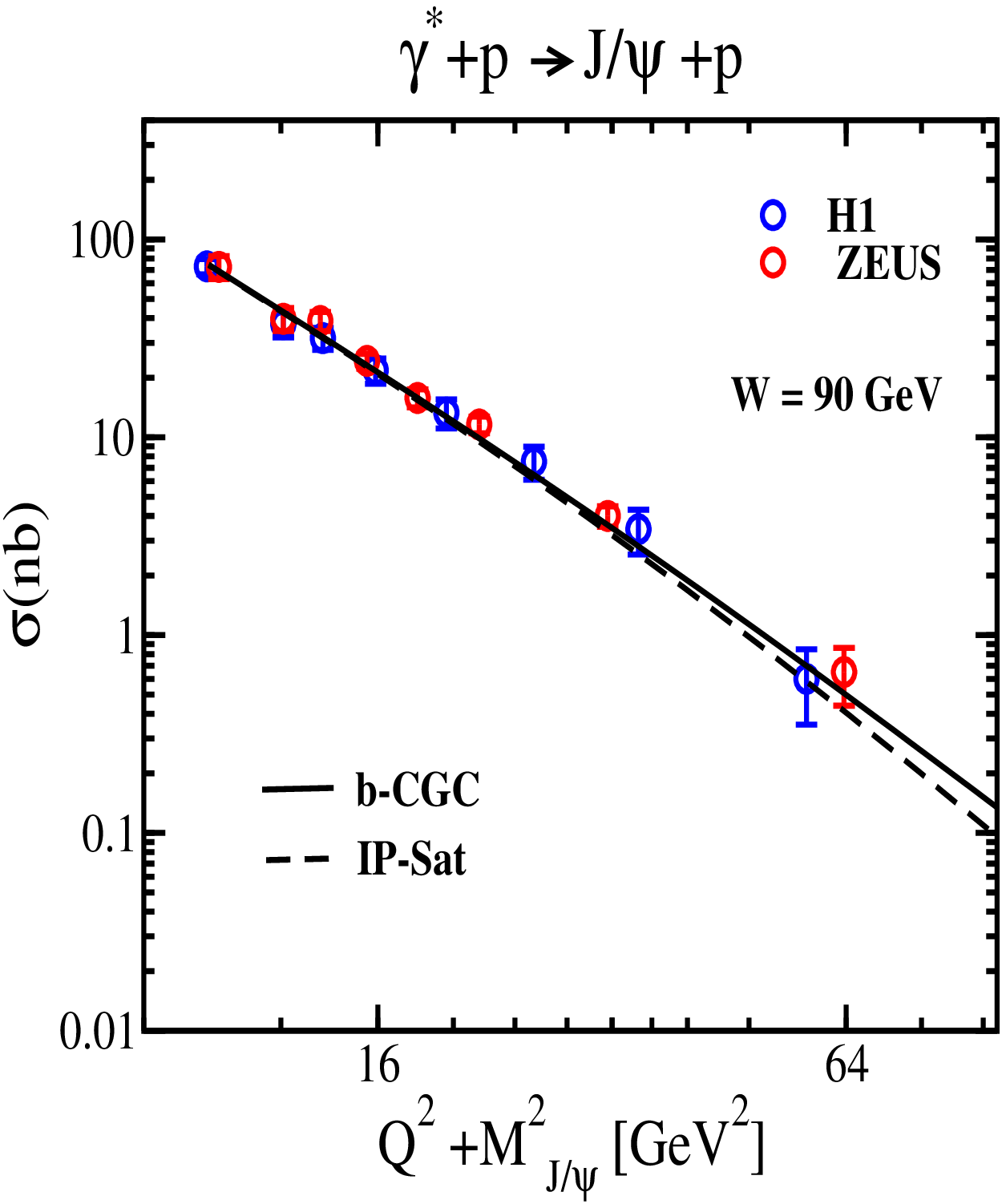}   
\includegraphics[width=0.4\textwidth,clip]{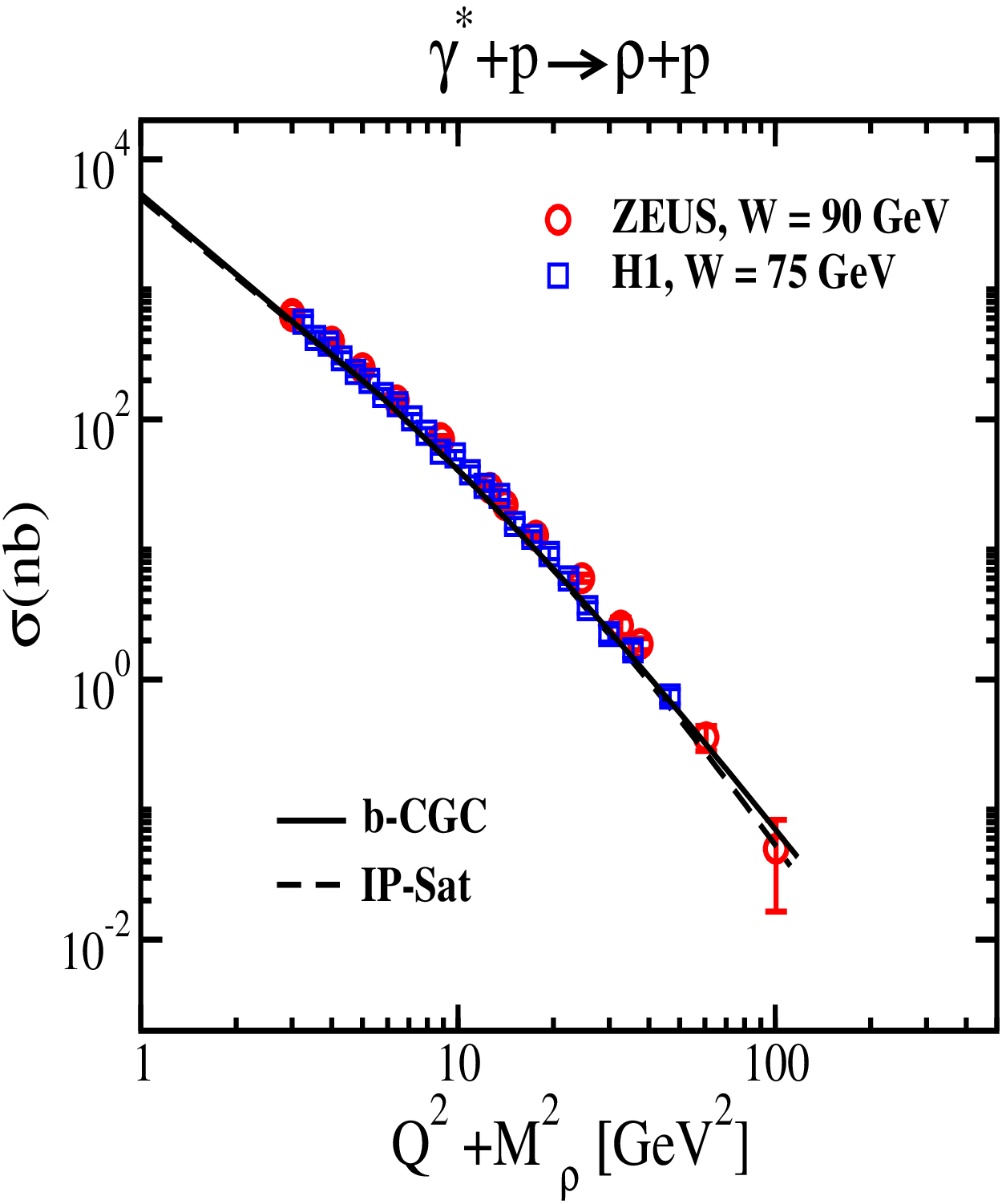}
\includegraphics[width=0.4\textwidth,clip]{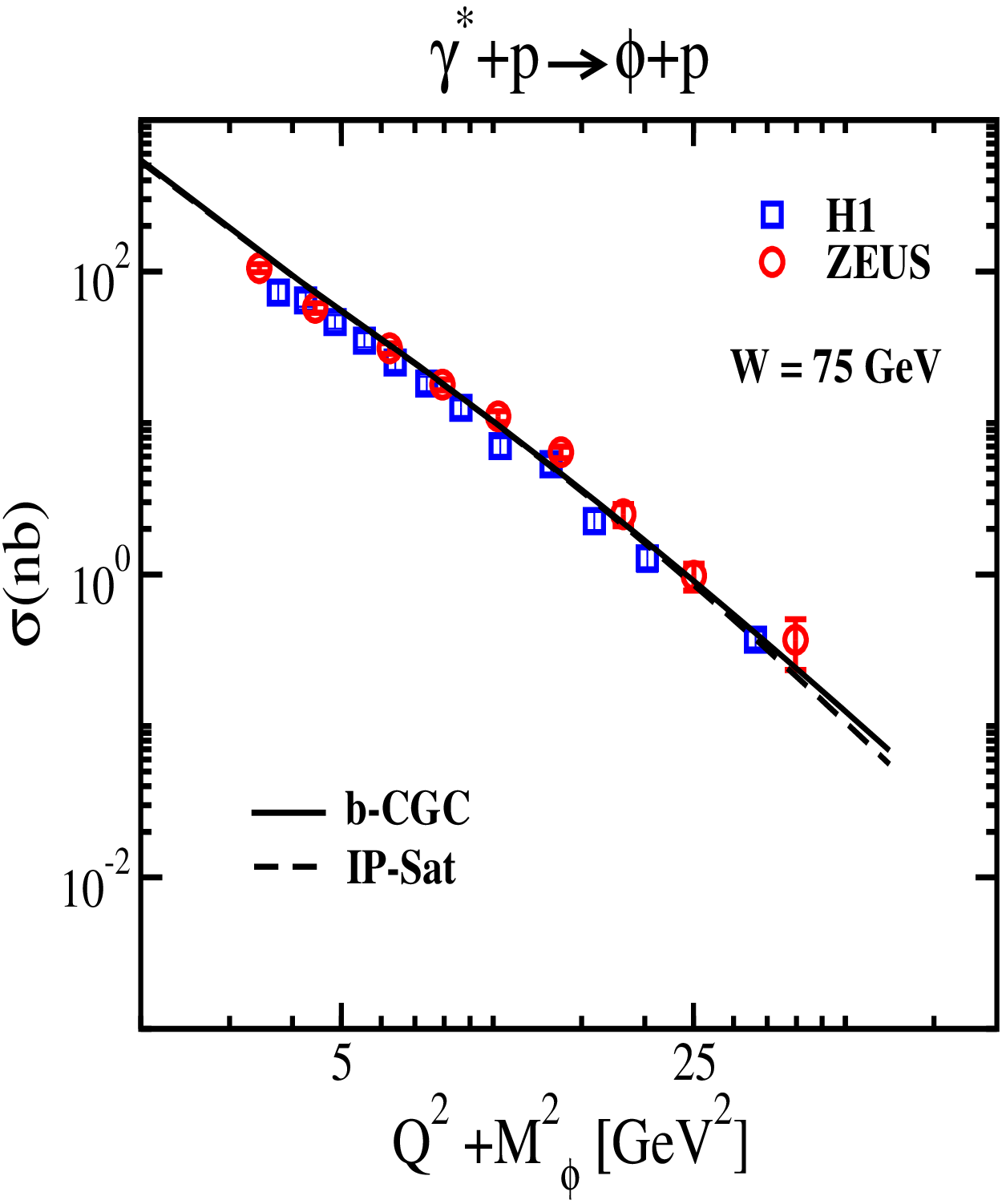}
\includegraphics[width=0.4\textwidth,clip]{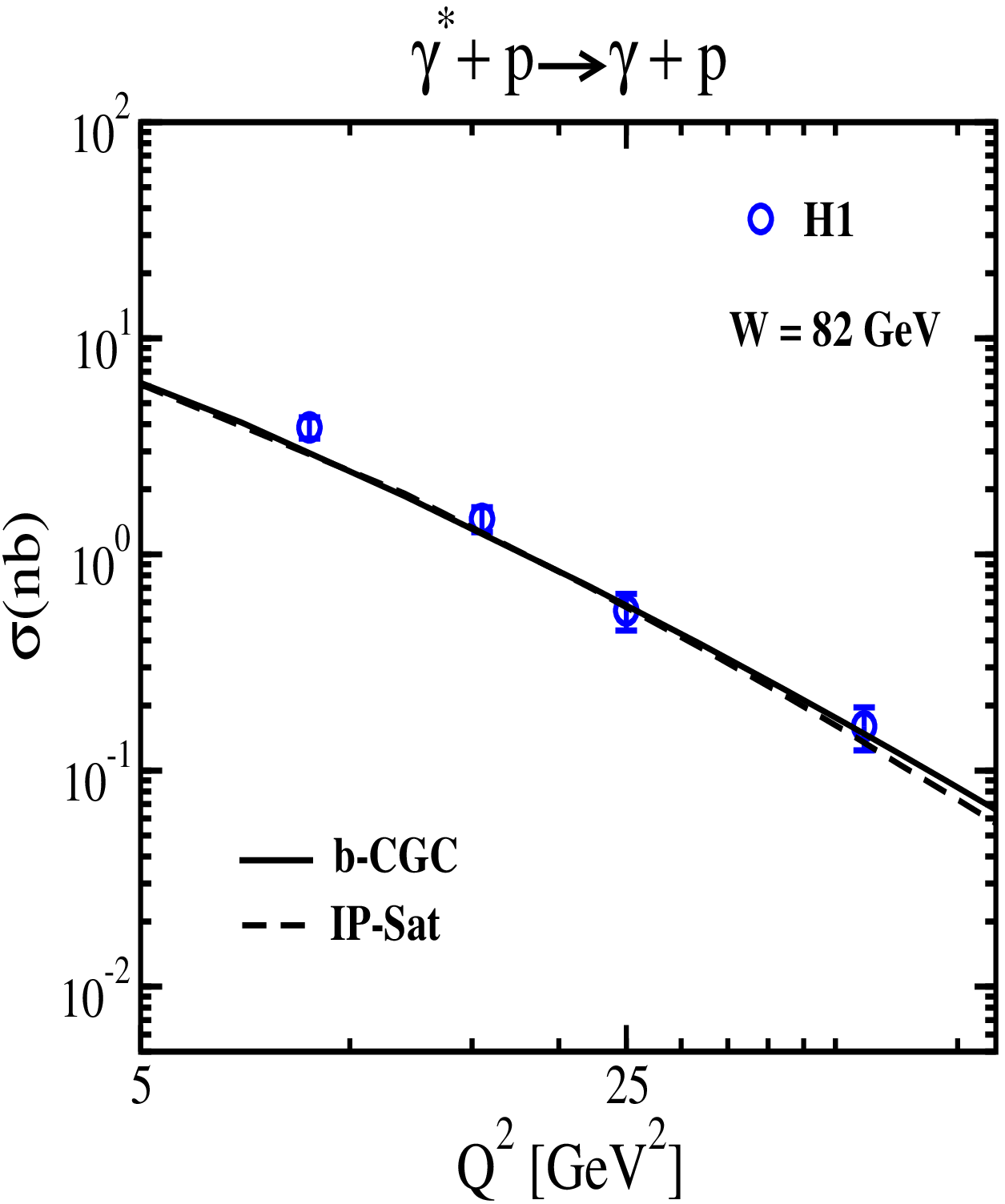}
\caption{Total vector meson cross-section $\sigma$ for $J/\Psi$, $\phi$, $\rho$ and DVCS, as a function of $Q^2+M^2_V$, compared to results from the b-CGC (solid lines) and the IP-Sat model (dashed lines), using parameter sets with $m_c=1.27$ GeV. The experimental data are from H1 and ZEUS collaborations \cite{Chekanov:2002xi,Chekanov:2004mw,Aktas:2005xu,Chekanov:2005cqa,Aaron:2009xp,Chekanov:2007zr,Aaron:2009ac,Chekanov:2008vy}.  }
  \label{f-vq}
\end{figure}     

\begin{figure}[t]       
\includegraphics[width=0.4\textwidth,clip]{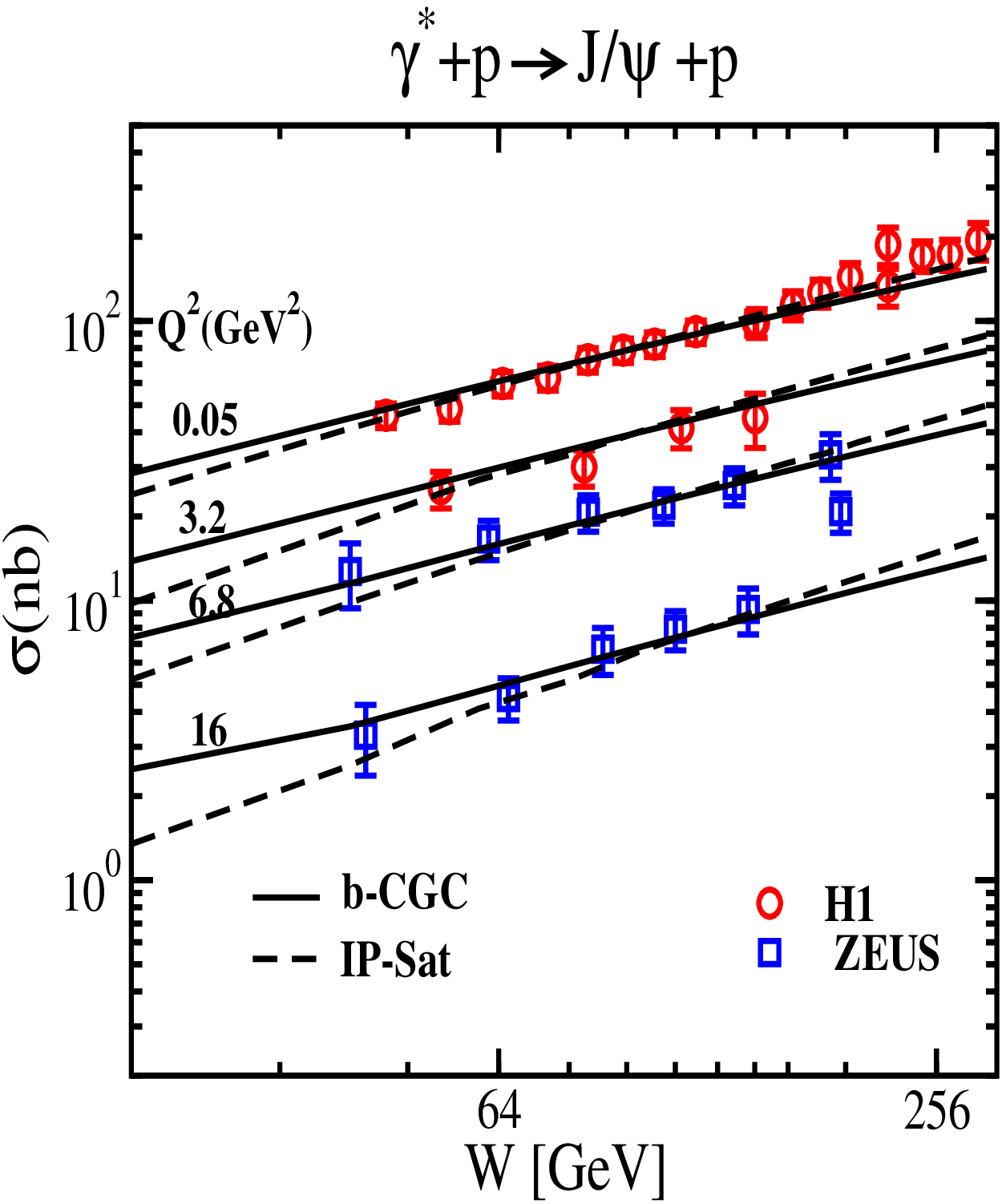}   
\includegraphics[width=0.4\textwidth,clip]{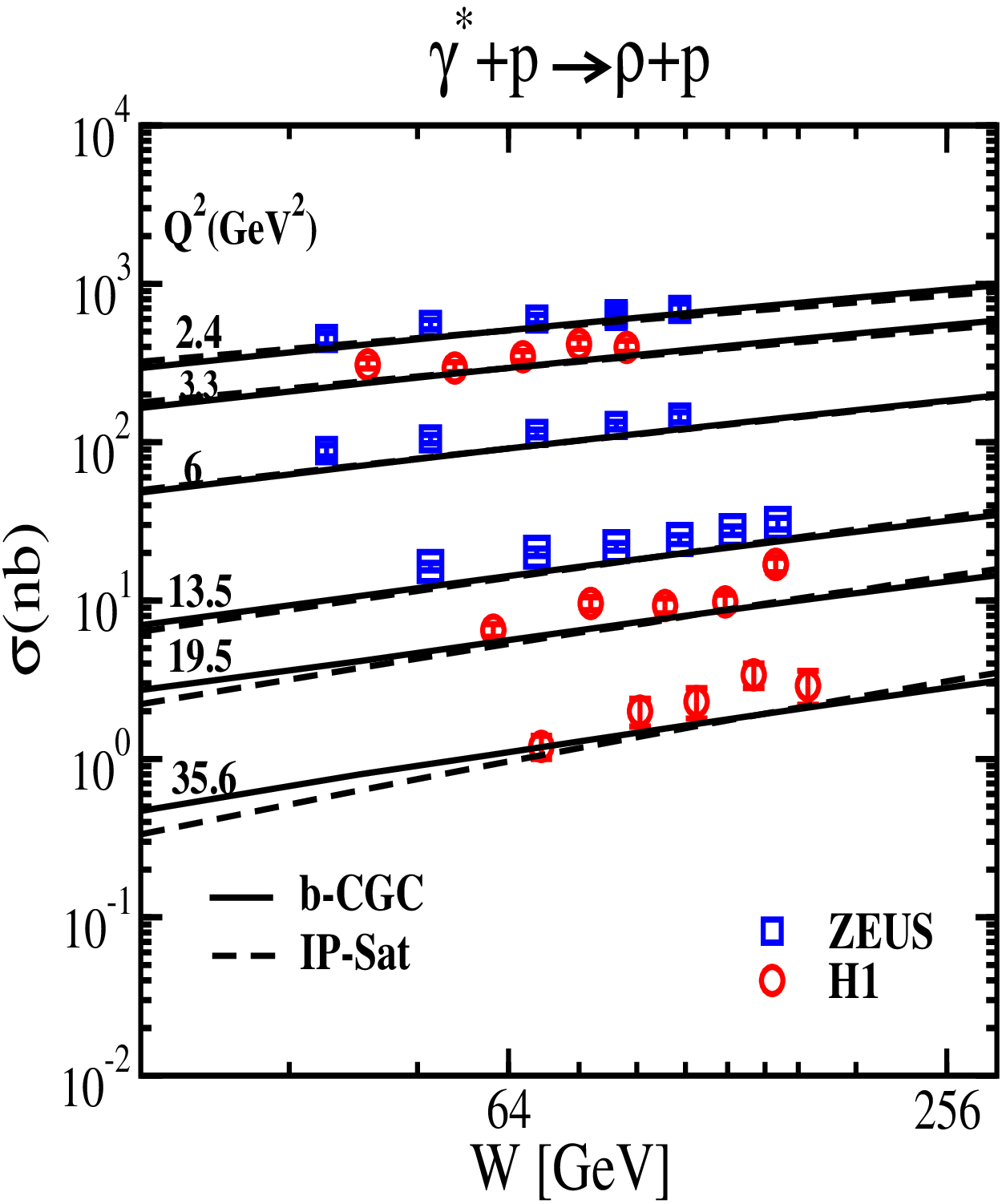}
\includegraphics[width=0.4\textwidth,clip]{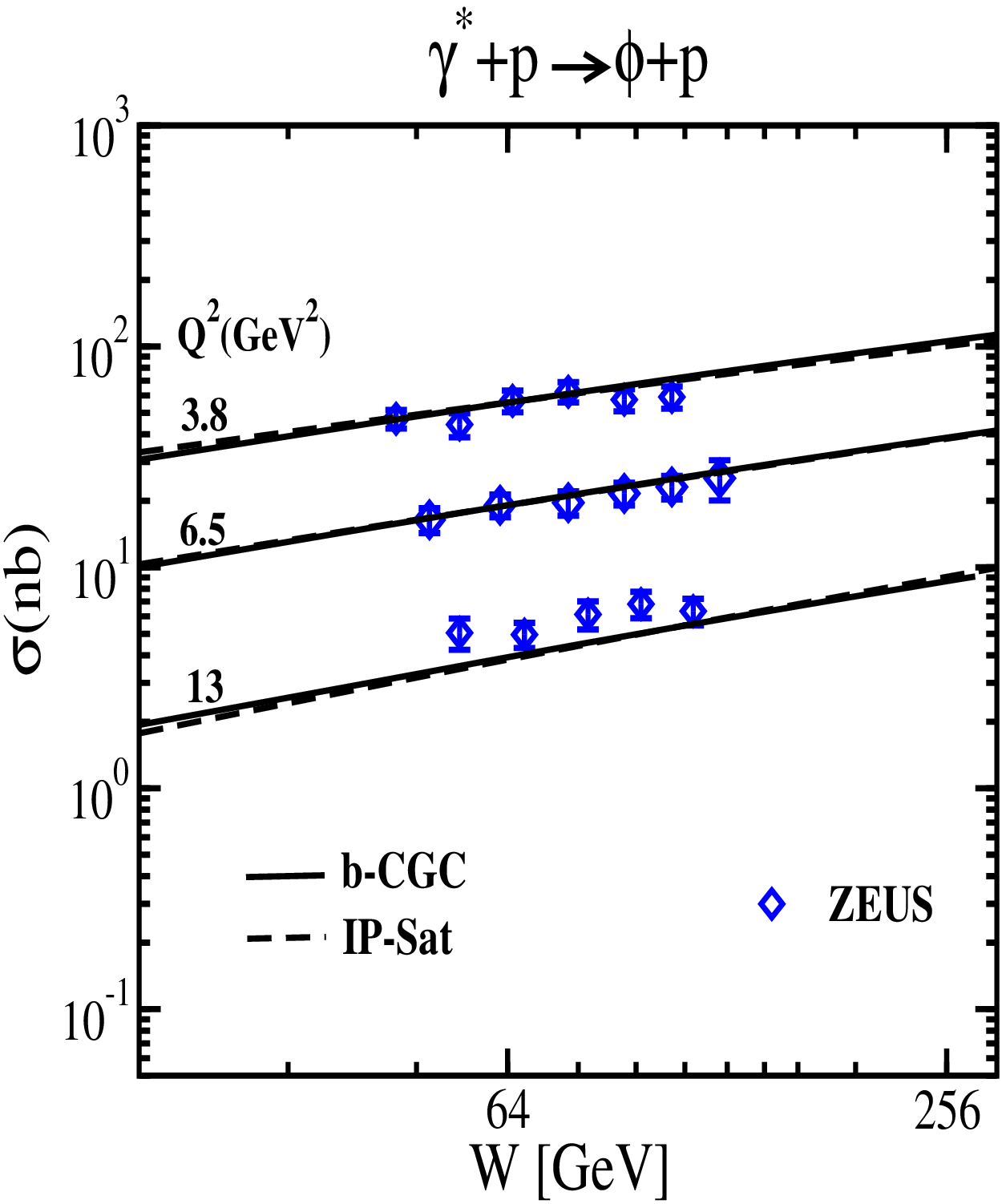}
\includegraphics[width=0.4\textwidth,clip]{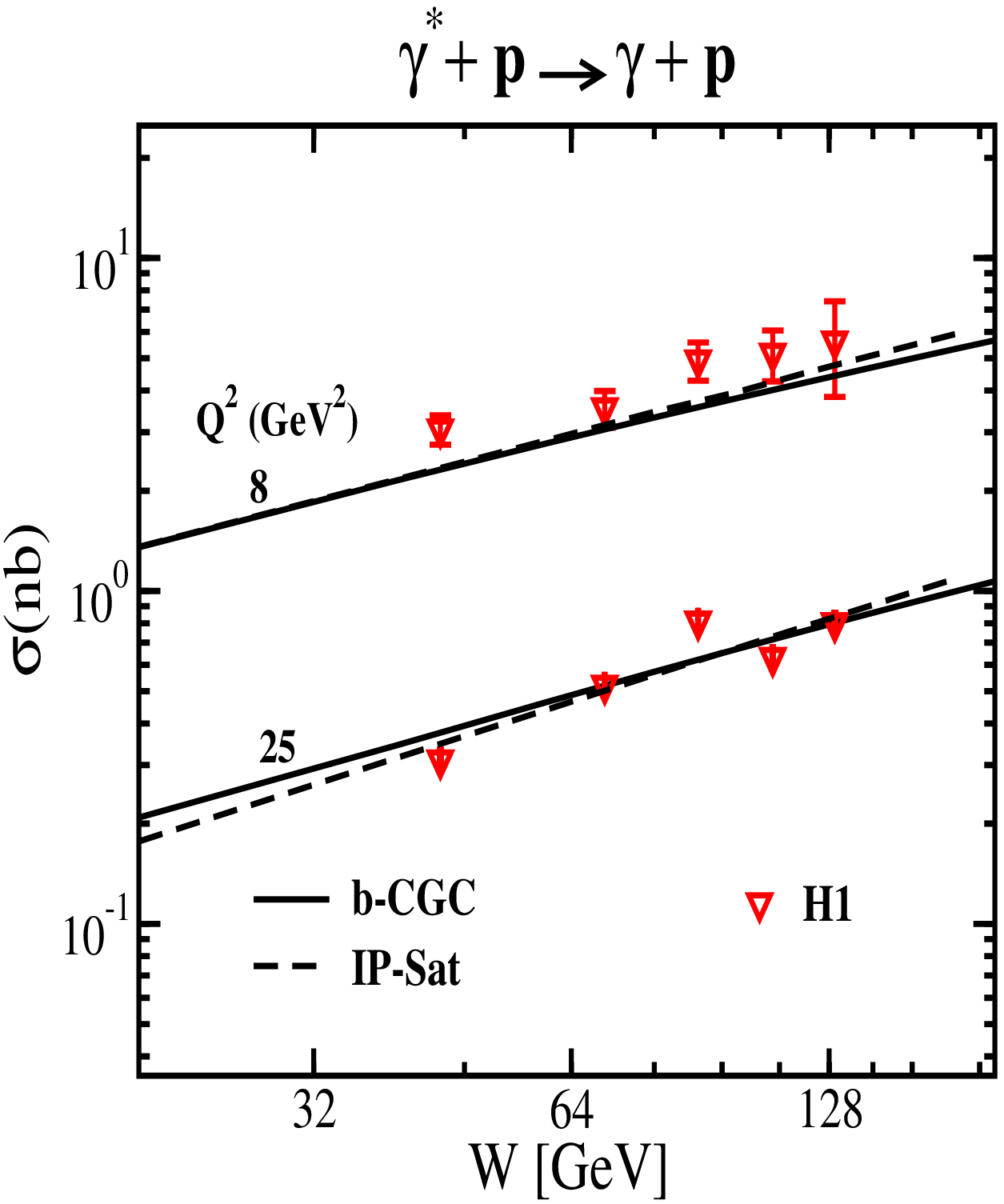}
\caption{Total vector meson cross-section and DVCS, as a function of $W$, compared to results from the b-CGC (solid lines) and the IP-Sat model (dashed lines), using parameter sets with $m_c=1.27$ GeV.
The experimental data are from H1 and ZEUS collaborations \cite{Chekanov:2002xi,Chekanov:2004mw,Aktas:2005xu,Chekanov:2005cqa,Aaron:2009xp,Chekanov:2007zr,Aaron:2009ac,Chekanov:2008vy}. }
  \label{f-vw}
\end{figure}

\begin{figure}[t]       
\includegraphics[width=0.32\textwidth,clip]{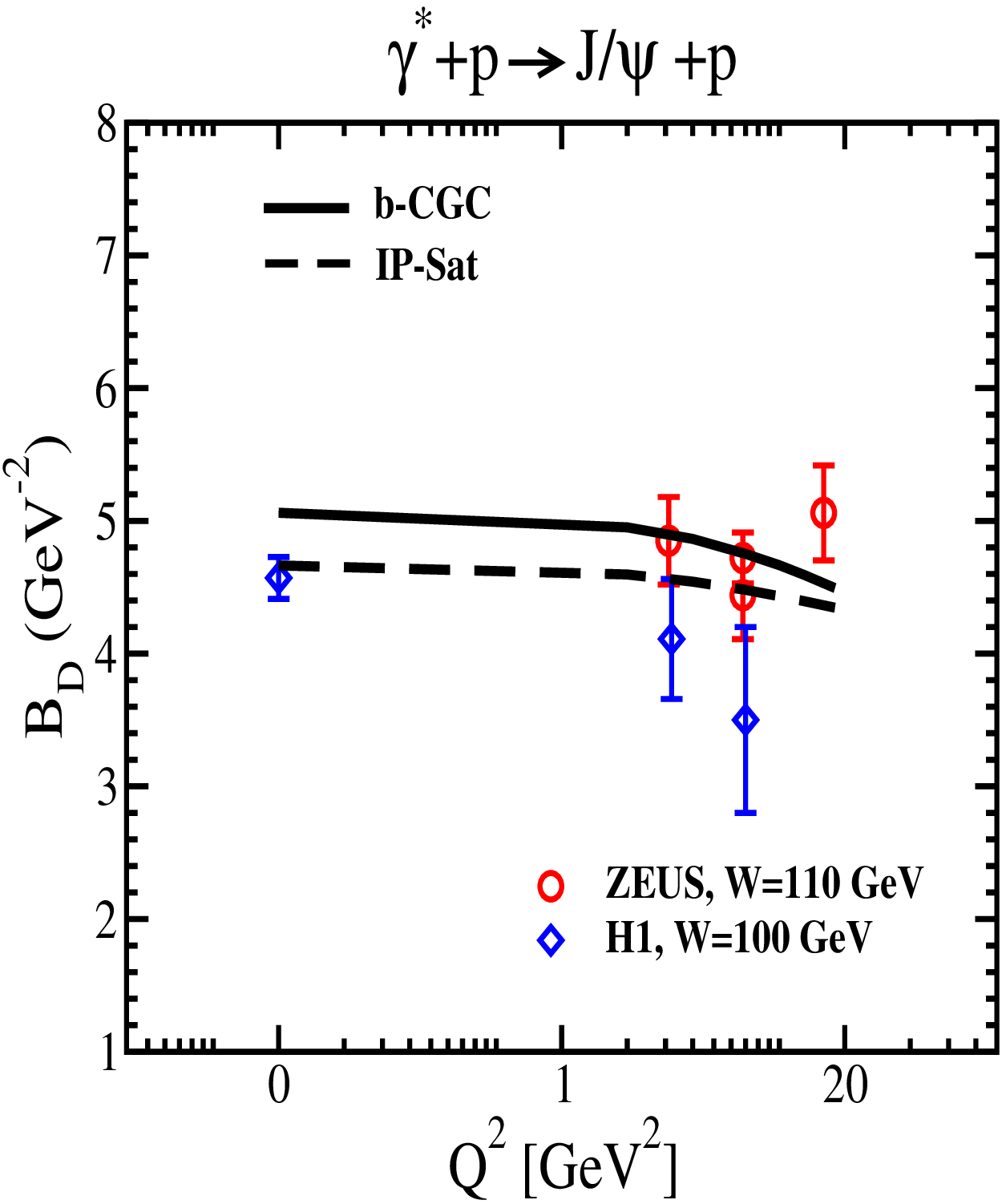}   
\includegraphics[width=0.32\textwidth,clip]{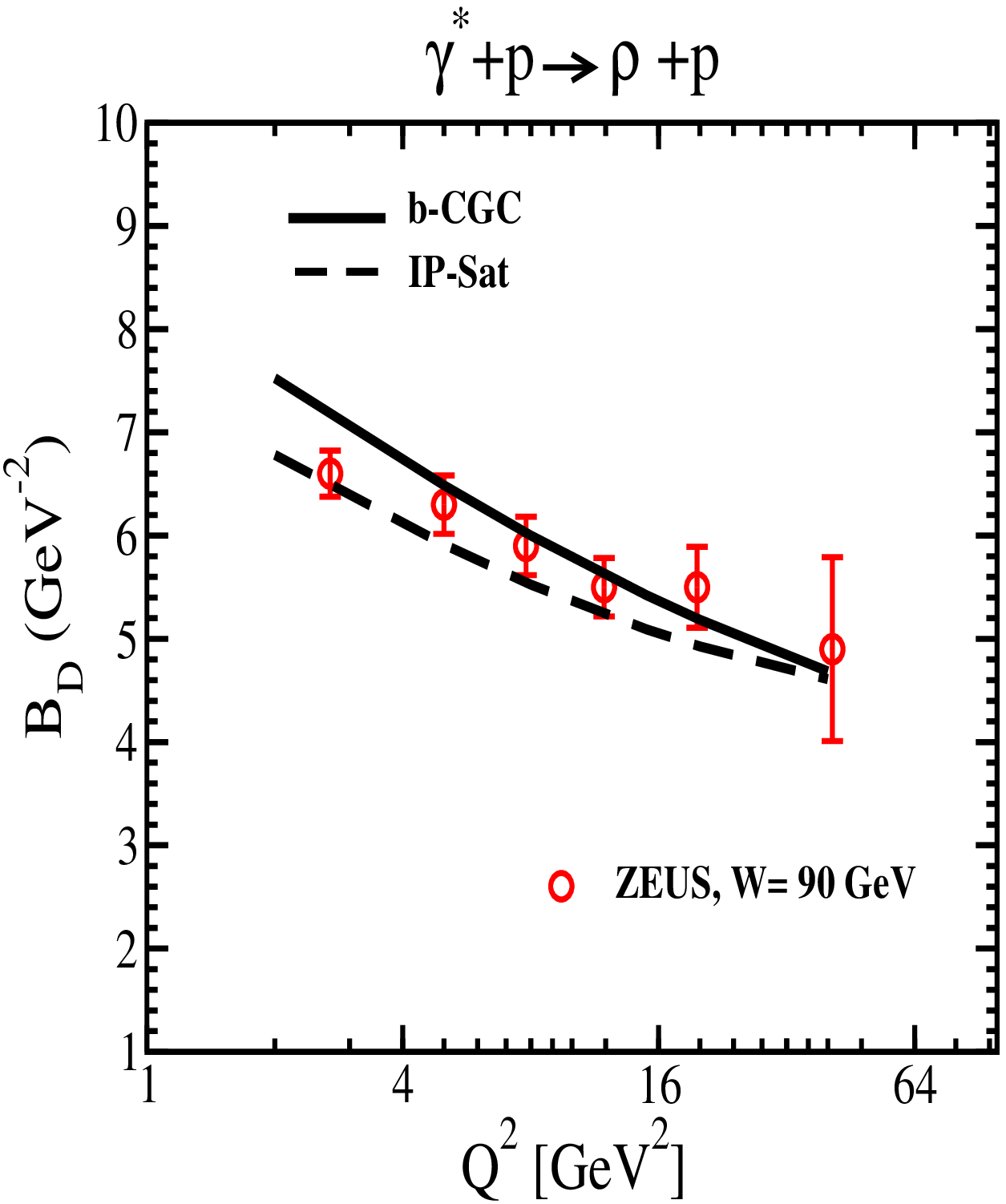}
\includegraphics[width=0.32\textwidth,clip]{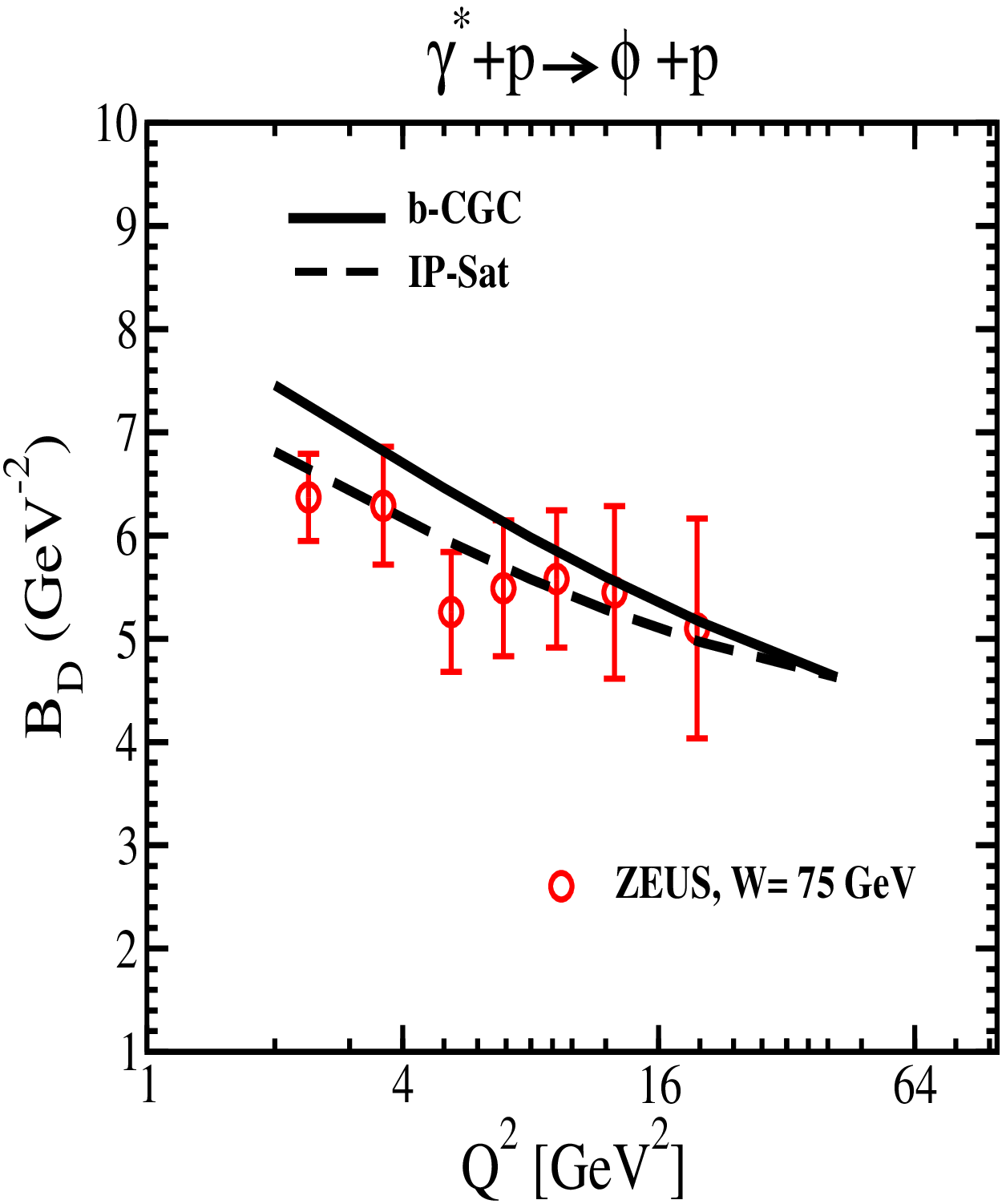}
 \caption{The slope $B_D$ of the t-distribution of exclusive vector-meson electroproduction, as a function of $Q^2$, in the b-CGC (solid lines) and the IP-Sat model (dashed lines). The experimental data are from \cite{Chekanov:2002xi,Chekanov:2004mw,Aktas:2005xu,Chekanov:2005cqa,Aaron:2009xp,Chekanov:2007zr}. }
  \label{f-bd-q}
\end{figure}  

\begin{figure}[t]       
\includegraphics[width=0.32\textwidth,clip]{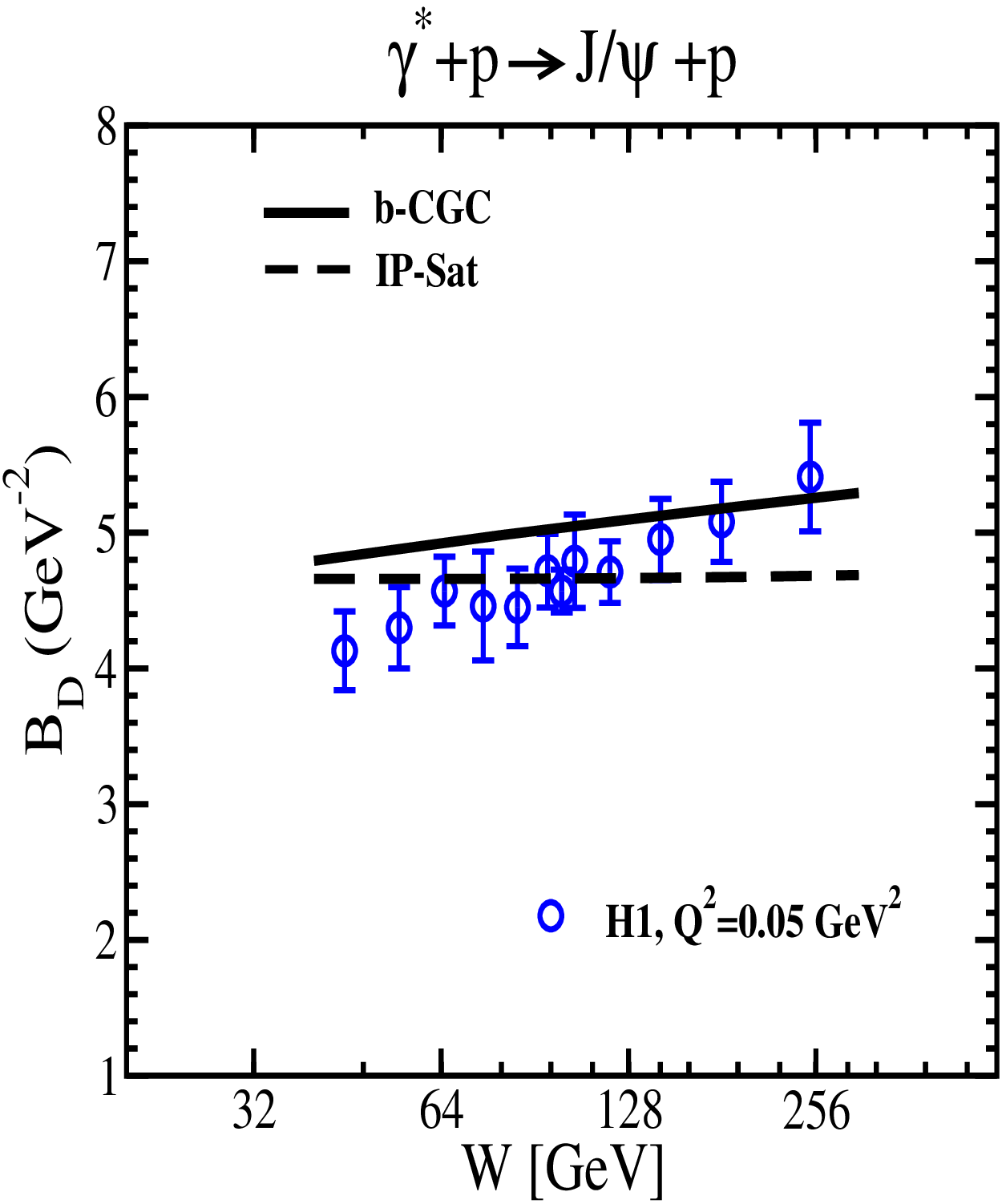}   
\includegraphics[width=0.32\textwidth,clip]{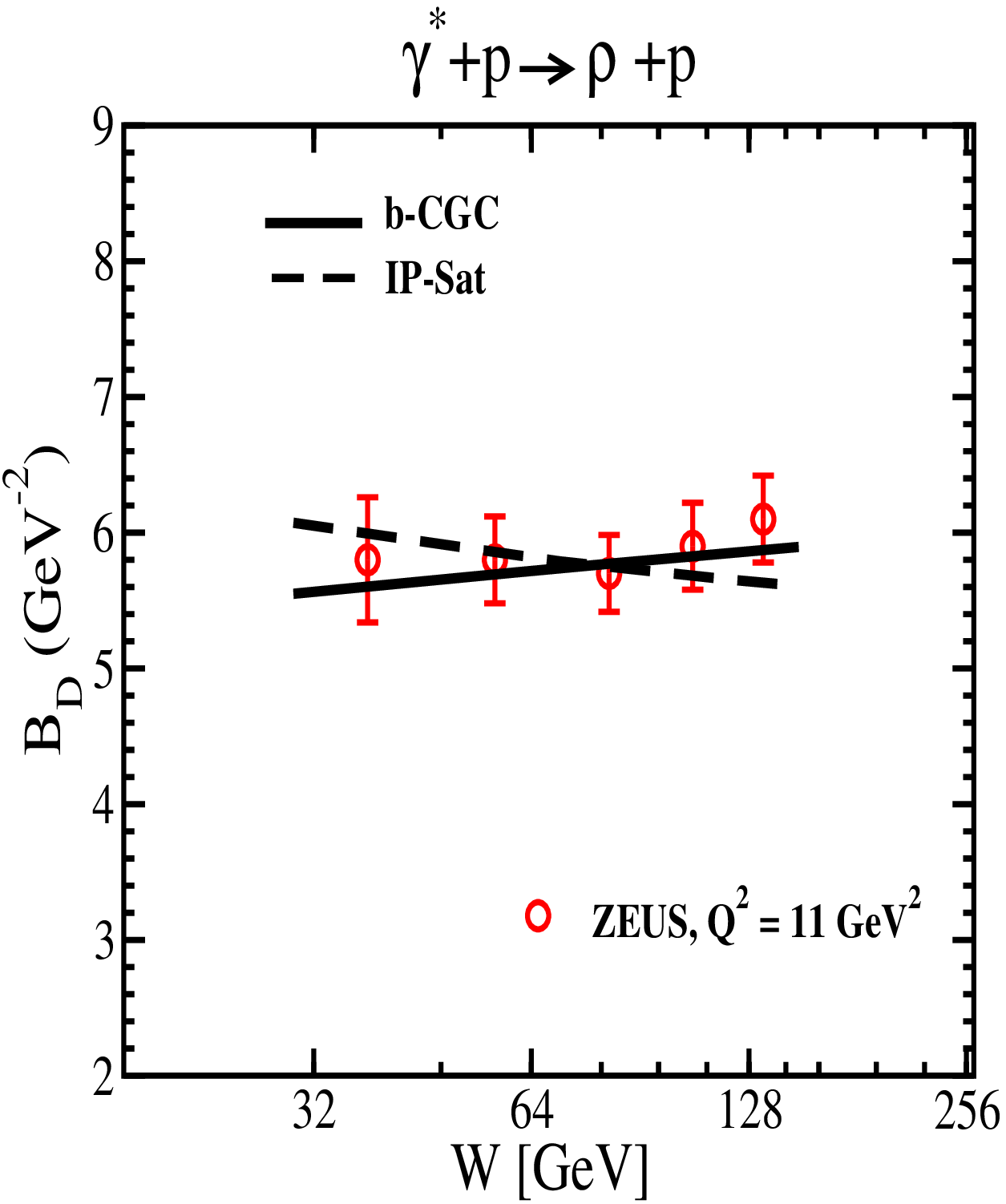}
\includegraphics[width=0.32\textwidth,clip]{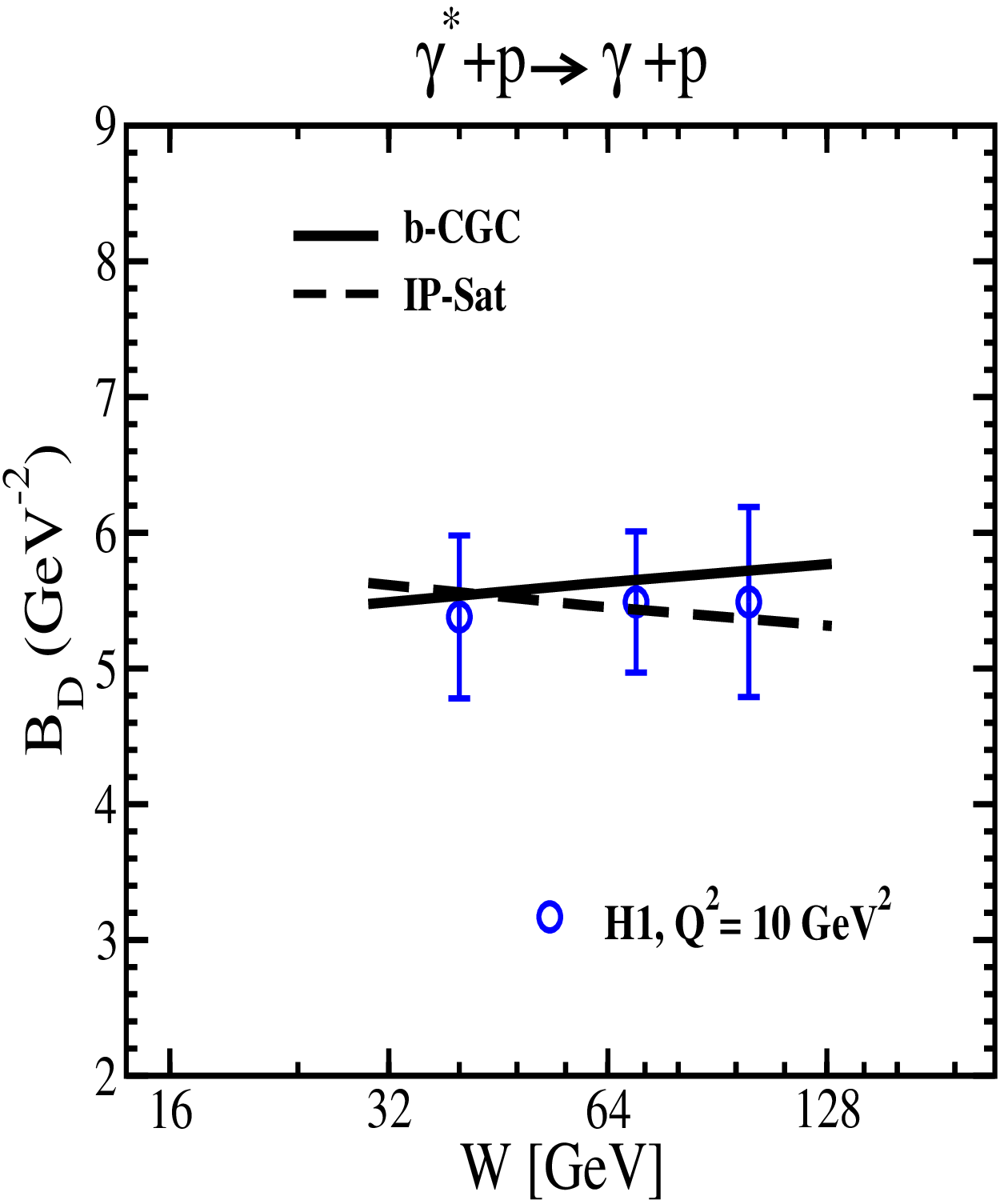}
 \caption{The slope $B_D$ of the t-distribution of exclusive vector-meson electroproduction and DVCS processes, as a function of $W$ in the b-CGC (solid lines) and the IP-Sat model (dashed lines). The experimental data are from \cite{Chekanov:2002xi,Chekanov:2004mw,Aktas:2005xu,Chekanov:2005cqa,Aaron:2009xp,Chekanov:2007zr,Aaron:2009ac,Chekanov:2008vy}. }
  \label{f-bd-w}
\end{figure}     


Next, we consider exclusive diffractive production at HERA in the b-CGC and the IP-Sat dipole model. We first focus on the $t$-distribution of  exclusive vector meson production and of Deeply Virtual Compton Scattering (DVCS). We fix the width of the impact-parameter profile $B_{CGC}$ of the saturation scale in \eq{qs-b} via a fit to the slope of the $t$-distribution of the diffractive $J/\Psi$ production, and find $B_{CGC}=5.5\,\text{GeV}^{-2}$. This value is smaller than what was obtained in Ref.\,\cite{watt-bcgc}. However, notice that there is a correlation between the two parameters $\lambda$ and $B_{CGC}$, and consequently the extracted value of $\lambda$ is larger here than in Ref.\,\cite{watt-bcgc}.  We recall that the parameter  $\lambda$ is mainly fixed via a fit to the reduced cross-section data of the  combined H1 and ZEUS collaborations.

In order to further test the b-CGC dipole model, in Figs.\,\ref{f-ch},\ref{f-vt},\ref{f-vq},\ref{f-vw}  we confront experimental data from H1 and ZEUS \cite{Chekanov:2002xi,Chekanov:2004mw,Aktas:2005xu,Chekanov:2005cqa,Aaron:2009xp,Chekanov:2007zr,Aaron:2009ac,Chekanov:2008vy,h1-jps} with model predictions for the $|t|$, $Q^2$ and $W$-dependence of the vector mesons  $J/\Psi$, $\phi$, $\rho$ and DVCS production, in various kinematics. 
 For the total cross-section, we performed the integral over $|t|$ up $1\,\text{GeV}^2$. 
Note that except for the $t$-distribution of $J/\Psi$ at a fixed $W$, namely a few data points in Fig.\,\ref{f-ch} left panel,  none of the data sets in the other figures are used into our fits, and the results of the model can be considered as predictions. It can be generally seen that the agreement between our results and  data is excellent.

Notice that both set of parameters in table \ref{t-2}, with charm masses $m_c=1.27, 1.4$ GeV, give very similar results for $\phi$, $\rho$ and DVCS production, while  $J/\Psi$ production cross-section is more sensitive to the charm quark mass at low $Q^2$ (see \fig{f-ch}). This is because the scale in the integrand of the cross-section is set by the charm mass for low virtualities $Q^2<m_f^2$, and thereby the cross-section becomes sensitive to the charm quark mass.  However, given rather large experimental error bars, there is no great preference between the two different parameter sets given in table \ref{t-2} for exclusive diffractive data (see  \fig{f-ch}). Nevertheless,  the proton structure functions and the data for $W$, $Q^2$ of $J/\Psi$ production are generally slightly better reproduced with the parameter set corresponding to lighter charm mass $m_c=1.27$ GeV, therefore in the rest of the paper for both the b-CGC and the IP-Sat model we will only show the results calculated by the parameter set with  $m_c=1.7$ GeV. In Figs.\,\ref{f-vt},\ref{f-vq} and \ref{f-vw} we compare our results in the b-CGC  (solid lines)  and the IP-Sat dipole (dashed lines) for exclusive diffractive vector mesons and DVCS production.

We see from \fig{f-vw} that the $W$-dependence of the cross-section follows a power-law behavior of the form $\sigma\propto W^\delta$, indicating the existence of the geometric scaling property in the diffractive data \cite{gs-dd}. Our extracted values of $\delta$, at different kinematics and for various vector mesons, are also in perfect agreement with experimental data from H1 and ZEUS \cite{Chekanov:2002xi,Chekanov:2004mw,Aktas:2005xu,Chekanov:2005cqa,Aaron:2009xp,Chekanov:2007zr}.

 In Figs.\,\ref{f-bd-q},\ref{f-bd-w} we show the $Q^2$ and  $W$ dependence of the slope of the $t$-distribution of exclusive vector-mesons electroproduction and DVCS, extracted from a fit of a form $d\sigma/d|t| \propto e^{-B_D t}$ within $|t|\in[0,1]\,\text{GeV}^2$, in the b-CGC and the IP-Sat dipole models.  As it is seen in Figs.\,\ref{f-bd-q},\ref{f-bd-w}, the experimental errors for the values of $B_D$ are rather large. This leads to some uncertainties in extracting the value of the parameter $B_{CGC}$, of about $0.5\,\text{GeV}^{-2}$. It is seen from \fig{f-bd-q} that the extracted values of $B_D$ at the same $Q^2$ are larger for lighter vector mesons than for $J/\Psi$ production, in accordance with data. This is due to the fact that the averaged dipole size, which dominates in the  wave function overlap, is different for light and heavy vector mesons, and it is mainly controlled by the inverse of $Q^2+M^2_V$. Therefore, at fixed virtuality the typical dipole size which participates in the interaction is bigger for lighter vector meson, and consequently it is expected that the same asymptotic behavior of the amplitude be maintained at higher virtualities for light vector mesons. The recent data from ZEUS \cite{Chekanov:2007zr} for diffractive vector meson production which cover a wider range of kinematics, indeed shows strong indication that at large $Q^2+M^2_V$ the value of $B_D$,  for various vector mesons, tends to saturate to a universal value, which in the dipole approach is mainly controlled by the impact-parameter profile of the saturation scale (or proton). This effect is also seen in \fig{f-bd-w}, where the slope of the $t$-distribution of $J/\Psi$ production seems to be approximately flat with respect to $Q^2$, in sharp contrast with light diffractive vector meson and DVCS production.  It is seen from \fig{f-bd-w} that the energy dependence of the slope $B_D$ at fixed $Q^2$, for different diffractive vector mesons production, is different in the b-CGC and the IP-Sat dipole models.  In the b-CGC model, the slope $B_D$ rises with $W$ at fixed $Q^2$, while we have almost the opposite trend in the IP-Sat model for light diffractive vector meson and DVCS production, and with little $W$ dependence for $J/\Psi$ production. This is mainly because the b-dependence of the dipole amplitude in the b-CGC is different from the IP-Sat model. In the b-CGC dipole model we have some non-trivial correlations between the effective impact-parameter profile of the proton and x, which leads to the $W$ dependence of the slope $B_D$.  Unfortunately,  the experimental data points in \fig{f-bd-w}  are limited in kinematics, with rather large error bars, and therefore at the moment the experimental data cannot conclusively discriminate between these two dipole models. The different extracted 
value of the slope $B_D$ in the b-CGC and the IP-Sat dipole models also leads to sizable effect in the  $t$-distribution of exclusive vector-mesons electroproduction and DVCS production at high $t$ (see \fig{f-vt}).

\section{Summary}
The new combined HERA data, which is significantly more precise than previous data, put tougher constrains on the model parameters of dipole models. 
In this paper we confronted the CGC and the b-CGC dipole models to the new combined data from HERA, and obtained the model parameters. The CGC and the b-CGC dipole model have only 4 and 5 free parameters, respectively. In the CGC dipole model, all free parameters are fixed by the DIS data, while in the b-CGC dipole model, 4 parameters are fixed via a fit to the DIS data and the last parameter, which determines the normalization of the total cross section for $\gamma^{*}p$ scatterings, is iteratively fixed via a fit to the slope of $t$-distribution of exclusive $J/\Psi$ photoproduction. The impact-parameter dependence of the b-CGC dipole model is crucial to have a unified description of  exclusive diffractive vector meson and DVCS, alongside DIS  processes. The b-CGC results were then compared to  the available data from HERA  for $F_2, F_2^{c\bar{c}}, F_L$, exclusive diffractive processes such as  $J/\Psi$, $\rho$, $\phi$ and the DVCS  production. Overall, the model provides an excellent description of data in the range $Q^2\in [0.75, 650]\,\text{GeV}^2$ and $x\le 0.01$. 

In the previous analysis of old HERA data in the b-CGC model \cite{watt-bcgc}, the extracted values of the anomalous dimension $\gamma_s=0.46$ and $\lambda=0.119$ \cite{watt-bcgc}  in the saturation scale were found to be significantly smaller than what may be expected from a perturbative calculation \cite{pqcd-s1,pqcd-s2,pqcd-s3,pqcd-s4,pqcd-s5,pqcd-s6,pqcd-s7}. The extracted values of $\gamma_s\approx 0.65$ and $\lambda \approx 0.20$ from the new combined HERA  (see table \ref{t-2}) are now approximately compatible with the perturbative expectation. Nevertheless, one should bear in mind that because of the impact-parameter dependence of the dipole amplitude, the b-CGC dipole model  intrinsically incorporates some non-perturbative physics which is beyond the weak-coupling approximation (see Refs.\,\cite{non-pqcd1,KR}). Other key features of our novel fit are the small values for the light quark masses, close to the current quark masses, and also the small value for the parameter $B_{CGC}$ in the impact-parameter profile of the saturation scale, compared to the old analysis.  

We compared our results obtained with the b-CGC dipole model to other well-known impact-parameter dependent saturation model, the so-called IP-Sat model \cite{ip-sat2}.  Both models include saturation physics. The b-CGC is based on the BK non-linear evolution, while the IP-Sat model incorporates the saturation effect via the Glauber-Mueller approximation, with the DGLAP evolution. We showed that most features of inclusive DIS and exclusive diffractive data, including the $Q^2$, $W$, $|t|$ and $x$ dependence, are correctly reproduced in both models.  Nevertheless, they give systematically different predictions beyond the current HERA kinematics: for the structure functions at very low $x$ and high virtualities $Q^2$ shown in Figs.\, \ref{f-f2}, \ref{f-f2c}, \ref{f-fl}, \ref{f-fl-h1} and for the exclusive vector meson and DVCS  production at  high $t$, shown in Figs.\,\ref{f-vt},\ref{f-bd-w}. The main differences between the b-CGC and the IP-Sat models can be traced back to the different power-law behavior of the saturation scale in $x$ and different impact-parameter dependence, see \fig{f-g2}, which leads to sizable effects at very low-x and large-t. Both models  give approximately similar saturation scale for proton $Q_S<1$ GeV in the HERA kinematics for $x<10^{-5}$, for the most relevant impact-parameter values $b\approx 1\div 4\,\text{GeV}^{-1}$.  It is also remarkable that
although the impact-parameter dependence in the b-CGC and the IP-Sat models are very different, both models lead to the same conclusion that the typical impact-parameter probed in the total $\gamma^{*}p$ cross-section is about  $b\approx 1\div 4\,\text{GeV}^{-1}$ with a median about $b\approx 2\div 3\,\text{GeV}^{-1}$ (see \fig{f-g3}). We stress that the $t$-distribution of all diffractive vector mesons, including $J/\psi$, $\phi$, $\rho$ as well as DVCS,  can be correctly reproduced by fixing  only one parameter: $B_{CGC}$ (describing the width of saturation scale in the impact-parameter space), despite the fact that the vector meson and DVCS wave functions are very different. This strongly hints at universality of the underlying dynamics and the importance of the impact-parameter dependence of the saturation scale in the proton.  

The b-CGC dipole model has been already quite successful in phenomenological applications at RHIC and the LHC. However, the parameters employed in these studies were determined from old HERA data, predating the combined data sets for the proton.  It will be of great interest to investigate the impact of the new fits on observables in proton-proton, proton-nucleus and nucleus-nucleus collisions.

It should be noted that in order to understand more rigorously the exact nature of the gluon saturation in the proton wave function in the DIS and diffractive processes, it is important to systematically investigate the effect of  higher order contributions beyond the current leading-log approximation. The photon impact-factor \cite{ipm} and the color dipoles evolution have been recently calculated to NLO accuracy \cite{bk-nlo}. Unfortunately, the full NLO calculations for exclusive diffractive processes in the dipole approach, namely the non-forward photon impact factor (and the non-forward photon wave function), with proper inclusion of the impact-parameter dependence of collisions and skewness effect \cite{ske}, are not yet available. These are important issues which are beyond the scope of this paper and certainly deserve separate studies.

\begin{acknowledgments}
We are very grateful to Stefan Schmitt for useful communication. This work is supported in part by Fondecyt grants 1110781 and 1100287. 
\end{acknowledgments}


\newpage

 \end{document}